\documentclass[aip,sd,groupedaddress,reprint,author-year,author-numerical]{revtex4-1}
\usepackage{natbib}
\usepackage{latexsym}
\usepackage{graphicx}
\usepackage{epstopdf}
\usepackage{amsmath}
\usepackage{amssymb}
\usepackage{multirow}

\begin{document}

\preprint{AIP/123-QED}
%Title of paper
\title{The nature of hydrogen in $\gamma$-alumina}

\author{Yunguo Li}
\email[]{Yunguo@kth.se}
\affiliation{Division of Materials Technology, Department of Materials Science and Engineering, KTH Royal Institute of Technology, SE-100 44 Stockholm, Sweden}
\author{Cl\'audio M. Lousada}
\email[]{cmlp@kth.se}
\affiliation{Division of Materials Technology, Department of Materials Science and Engineering, KTH Royal Institute of Technology, SE-100 44 Stockholm, Sweden}
\author{Pavel A. Korzhavyi}
\email[]{pavelk@kth.se}
\affiliation{Division of Materials Technology, Department of Materials Science and Engineering, KTH Royal Institute of Technology, SE-100 44 Stockholm, Sweden}

\homepage[]{}
%\thanks{}

\altaffiliation{}

\date{\today}

\begin{abstract}
Gibbs free energy models are derived from the calculated electronic and phonon structure of two possible models of $\gamma$-alumina, a defective spinel phase and a hydrogenated spinel phase. The intrinsic vacancies and hydrogen in the two structural models give rise to a considerable configurational (residual) entropy and significantly contribute to thermodynamic stability and physical-chemical properties of $\gamma$-alumina, which was neglected in previous studies but considered in this work. The electronic densities of states, calculated using a hybrid functional for the two structural models of $\gamma$-alumina, are presented. The dynamic stability of the two phases is confirmed by full-spectrum phonon calculations. The two phases share many similarities in their electronic structure, but can be distinguished by their vibrational spectra and specific heat. The defective spinel is found to be the ground state of $\gamma$-alumina, while the hydrogenated spinel to be a metastable phase. However, dehydration of the metastable phase into the ground state is expected to be slow due to the low diffusion rate of H, which leaves hydrogen as a locked-in impurity in $\gamma$-alumina.
%%%%
\end{abstract}

% insert suggested PACS numbers in braces on next line

% insert suggested keywords - APS authors don't need to do this
%\keywords{}

%\maketitle must follow title, authors, abstract, \pacs, and \keywords
\maketitle

%%%%%%%%%%%%%%%%%%%%%%%%%%%%%%%%%%%%%%%%%%%%%%%%%%%%%%
% body of paper here - Use proper section commands
% References should be done using the \cite, \ref, and \label commands

\section{Introduction\label{I}}
Al$_2$O$_3$ also known as alumina is an important material due to its diverse industrial and technological applications and also due to its role in various geochemical processes.\cite{dorre1984alumina,hart1990alumina} Besides its most stable form, $\alpha$-alumina, there are also several transition aluminas that can be obtained by the thermal decomposition of aluminum hydroxides or oxyhydroxides.\cite{zhou1991, jace1995} The $\gamma$-alumina phase has been extensively studied mostly due to its applications as a catalyst or catalyst support.\cite{catalrev1978, hellman2008activation, scott1987molecular, wang2004dopants}
However, in spite of several experimental and theoretical works that have been devoted to the determination of its exact structure, the structure of $\gamma$-alumina is still not fully understood.\cite{stumpf1950thermal, zhou1991, catalrev1978, ushakov1984structure, tsyganenko1990infrared, sohlberg1999hydrogen, wolverton2000phase, PhysRevB.78.014106, paglia2005determination, PhysRevB.69.041405, ferreira2011direct} The inherent poor crystallinity of $\gamma$-alumina, that can be prepared through various synthetic routes, significantly increases the difficulty of the problem. Besides, the possible presence of hydrogen in $\gamma$-alumina is still not understood.\cite{zhou1991, catalrev1978, ushakov1984structure, tsyganenko1990infrared, sohlberg1999hydrogen, wolverton2000phase, PhysRevB.78.014106, paglia2005determination, PhysRevB.69.041405, ferreira2011direct} The clarification of the structural and physical-chemical properties of $\gamma$-alumina helps to understand the catalytic properties of this material. The catalytic properties of $\gamma$-alumina are related with its Lewis acidic surface due to the presence of hydroxyl groups.\cite{trueba2005gamma,digne2004use} This is because water molecules can undergo dissociative adsorption at the surface of $\gamma$-alumina.\cite{trueba2005gamma} However, this feature alone cannot be responsible for the differences in catalytic properties of $\gamma$-alumina and $\alpha$-alumina; because hydroxylated $\alpha$-alumina also has Lewis acidic surface sites due to dissociative water adsorption.\cite{hass1998chemistry} The hydrogenation of $\gamma$-alumina was suggested to be the possible reason for the catalytic differences between this and the other aluminas.\cite{dowden195056} The cation vacancies of $\gamma$-alumina are possible traps for protons, and it was found that the penetration of protons from the surface into the Al vacancies of the subsurface is very likely to occur and the energy barrier for this process is 1.05 eV/proton.\cite{rashkeev2007hydrogen, sohlberg1999hydrogen}

There are mainly two points of view in what concerns the structure of $\gamma$-alumina. These points of view differ mostly with regard to the content of hydrogen in the structure. One point of view states that $\gamma$-alumina is a stoichiometric transition alumina with a defective spinel structure. This model has its origins in  X-ray diffraction studies and dates back to 1935.\cite{lippens1970} The proposed structure is closely related to the MgAl$_2$O$_4$ magnesium aluminate spinel structure which contains tetrahedral cation sites (Mg sites) and octahedral cation sites (Al sites). The cubic unit cell of this magnesium aluminum spinel comprises 8 Mg cations (at 8a sites), 16 Al cations (at 16d sites) and 32 close-packed O anions (at 32e sites). The cation sites are not fully occupied in order to meet the stoichiometry of $\gamma$-alumina, and 2$\frac{2}{3}$ cation sites per cell have to be vacant. Conventionally, the formula is designated as $\square_{2\frac{2}{3}}$Al$_{21\frac{1}{3}}$O$_{32}$, where $\square$ stands for the vacancy. It is still a matter of debate which cation sites are preferred by the vacancies. It has been reported that the vacancies are formed preferably at the octahedral sites, but other observations indicate the preference for tetrahedral sites.\cite{lippens1970, sohlberg2000bulk, paglia2004boehmite}  The fraction of tetrahedrally coordinated Al should be limited between 25\% and 37.5\%. Early experiments by Verwey and Jagodszinski suggested the presence of vacancies at octahedral positions, with tetrahedrally coordinated Al amounting to 37.5\%.\cite{verwey1935electrolytic, jagodzinski1958, lippens1970, sohlberg2000bulk} However, a later experimental work by Saalfeld reported a different result with tetrahedrally coordinated Al amounting to 25\%.\cite{saalfeld1960strukturen, lippens1970} The fraction of tetrahedrally coordinated Al in $\gamma$-alumina determined by solid-state NMR is around 21$\sim$31\%.\cite{lee1997distribution, pecharroman1999thermal, john1983characterization} A computational study based on density functional theory (DFT) by Mo \emph{et al.}, suggested that the octahedral sites are more favorable for the formation of vacancies by 3.7 eV/vacancy.\cite{mo1997electronic} Additionally, other studies point towards the preference for vacancy formation at octahedral sites. In a recent work, Sun \emph{et al.} performed DFT calculations and Rietveld simulations to confirm the energetic preference for vacancies at octahedral sites.\cite{sun2006examination} Nevertheless, it has also been shown that the defective spinel structure with cation vacancies distributed between the two types of sites is in better agreement with experimental X-ray powder diffraction pattern.\cite{sun2006examination, paglia2006comment, digne2006comment}  Overall there are discrepancies between theoretical and experimental results in what concerns the location of vacancies. The results of Kn\"ozinger and Ratnasamy have shown a considerably disordered occupancy of tetrahedrally coordinated Al. This would make $\gamma$-alumina similar to $\eta$-alumina.\cite{catalrev1978} Zhou and Snyder have also shown evidence for disordered occupancy of tetrahedrally coordinated Al.\cite{zhou1991} The classical molecular dynamics simulation based on pairwise additive empirical potential functions performed by Alvarez confirmed the occupancy of non-spinel sites.\cite{alvarez1992molecular} Overall, the computational data point towards the preference for vacancies at octahedral sites. The disorder has its origins in temperature effects and synthetic factors.

Another type of suggested structures are the hydrogenated models of $\gamma$-alumina. The first model was introduced by Dowden in 1950.\cite{dowden195056} It was proposed that, in the presence of water, hydrogen can occupy the cation vacancies in $\gamma$-alumina. Later, de Boer and Houben conducted an X-ray diffraction (XRD) study and suggested a hydrogenated spinel structure denoted as Al$_2$O$_3 \cdot n$H$_2$O to account for the mass balance by variable water content $n < 1$.\cite{de1952proceedings} Such notation often leads people to take $\gamma$-alumina as a hydrated crystal. In the cited work, it was also suggested that the stoichiometry Al$_2$O$_3 \cdot$(1/5)H$_2$O with $n = 1/5$ corresponds to a perfect hydrogenated spinel HAl$_5$O$_8$. This conclusion was based on the analysis of several types of $\gamma$-alumina crystals with different hydrogen content.\cite{de1952proceedings} The presence of hydrogen in bulk $\gamma$-alumina was independently observed by Maciver and L\'enoard.\cite{maciver1963catalytic, leonard1969structure} In a NMR investigation by Pearson, the measured fraction of hydrogen in the bulk (rather than on the surface) was $n=0.18$ which very close to the value $n=0.2$ of the perfect hydrogenated spinel structure.\cite{pearson1971wide} Another hydrogenated spinel model proposed by Soled consisted in replacing O$^{2-}$ by two surface OH$^-$ groups.\cite{Soled1983252} However, the growth of $\gamma$-alumina by ion implantation and the annealing of $\alpha$-alumina ruled out this assumption.\cite{white1989ion, yu1995high} Other studies using diverse techniques have pointed towards the existence of hydrogenated structures. Using the XRD technique, Ushakov and Moroz found that only the structure containing residual bulk hydrogen could account for the experimentally recorded XRD patterns.\cite{ushakov1984structure} Tsyganenko \emph{et al.} observed OH$^-$ peaks in the infrared spectrum which were assigned to bulk OH$^-$ rather than surface OH$^-$ groups.\cite{tsyganenko1990infrared} In recent studies, DFT calculations were used to investigate the role of hydrogen in bulk $\gamma$-alumina. Sohlberg \emph{et al.} found that the hydrogenated spinel structures described as H$_{3m}$Al$_{2-m}$O$_3$ are considerably lower in energy than the defective spinel structures (with the perfect hydrogenated spinel structure HAl$_5$O$_8$ to be the lowest in energy). Furthermore, the calculated vibrational frequencies of bulk OH bonds agree well with experiments.\cite{sohlberg1999hydrogen} However, the calculated volume of $\alpha$-alumina has a large deviation from experimental data, which raises questions about the accuracy of the calculations. In a study by Wolverton and Hass, it has been shown that the hydrogenated spinel structure (HAl$_5$O$_8$) is thermodynamically unstable with respect to the decomposition into Boehmite and defective spinel structure of $\gamma$-alumina.\cite{wolverton2000phase} Yet, the studies of Sun \emph{et al.} suggest that the hydrogenated spinel structure is more stable than the defective spinel structure at $T<550$~K, and the opposite for $T>550$~K.\cite{sun2006examination} 

To enable for a careful analysis of relative stabilities for the competing structures it is necessary to make the expression for the Gibbs free energy as complete as possible. Worth of notice is the fact that, none of the previous \textit{ab initio} studies of $\gamma$-alumina has taken into consideration the vibrational contribution to the Gibbs free energy. The vibrational and thermal properties of alumina are of great interest to many industrial processes. Data on the thermal properties of $\gamma$-alumina are scarce due to its structural complexity.

Furthermore, configurational entropy was never considered and evaluated in previous \textit{ab initio} studies of this system. A part of configurational entropy associated with frozen-in disorder, referred to as residual entropy, originates from the degeneracy or near degeneracy in energy of a manifold of atomic configurations in the structures of some aluminum oxides and hydroxides. The disordered distribution of vacancies and hydrogen may be present in these compounds even at low temperatures and give rise to significant residual entropy contributions which may be important for the determination of relative phase stabilities. 

In this work, we perform DFT calculations of the electron and phonon spectra for the two structural models of $\gamma$-alumina as well as for Boehmite AlOOH. The residual entropy of these phases is evaluated using the approach proposed by Pauling.\cite{pauling1935structure} Using the calculated data, the vibrational and thermal properties of the compounds are calculated, including the heat capacity, entropy, and Gibbs free energy. We also present the calculated phonon dispersions and Raman-active modes for the defective spinel $\gamma$-alumina, hydrogenated spinel $\gamma$-alumina, and Boehmite.

\section{Methodology\label{II}}
\subsection{Electronic structure and phonon spectrum calculations\label{II.a}}
The present calculations are based on density functional theory (DFT) and use a plane-wave basis set, as implemented in the Vienna Ab initio Simulation Package (VASP).\cite{kresse1999ultrasoft, blochl1994projector} The interaction between the ions and valence electrons is described by the projector augmented wave (PAW) method.\cite{kresse1996efficiency, kresse1996efficient, kresse1993ab} Most of the calculations, including structural relaxations and phonon spectra calculations, are done on the level of generalized gradient approximation (GGA), employing the exchange-correlation functional by Perdew, Burke and Ernzerhof (PBE).\cite{perdew1996generalized} In general, semilocal GGA functionals are known to underestimate the electronic bandgap of semiconductors and insulators, while providing good structural accuracy.\cite{PhysRevB.77.165107, C3CY00207A} 

To derive accurate Gibbs free energy of the system, one needs a functional that provides satisfactory structural accuracy.
It has been shown that the GGA based functionals implemented in VASP can provide accurate structure and energy data for aluminum oxides and oxyhydroxides.\cite{wolverton2000phase, demichelis2010} Therefore, the relative thermal stabilities of the materials studied in this work are assessed on the basis of PBE calculations. However, we also employ the hybrid functional approach to compare with the PBE results and assure the quality of calculations. For that purpose we use PBE0 functional \cite{adamo:6158} which contains 25\% of the exact nonlocal Hatree-Fock (HF) exchange, and 75\% of the PBE exchange, and 100\% of the PBE correlation energy.

In our work, a plane-wave cut-off energy of 550 eV was enough to reach convergence and was used in the calculations of defective spinel $\gamma$-alumina. For the PBE type of calculations, a k-mesh of 3$\times$9$\times$9 was tested to reach energy convergence, and 1$\times$3$\times$3 was used for the PBE0 type of calculations. The total energy difference was converged to less than 1$\times$10$^{-8}$ eV/unit cell, and the force acting on each atom is converged within $10^{-4}$ eV/\AA. Hydrogenated spinel $\gamma$-alumina was also considered in our study, for comparison with the defective spinel model. A k-mesh of 9$\times$9$\times$9 and a plane-wave cut-off energy of 550 eV were used for the PBE type of calculations of this structure, and a 3$\times$3$\times$3 mesh was used for the PBE0 type of calculations. The total energy and force were converged with the same accuracy as for the defective spinel structure. 

Although the location of the vacancies is not yet fully understood from experimental data, previous theoretical studies have pointed towards the preference for formation of vacancies at octahedral sites.\cite{mo1997electronic, sun2006examination}  In Ref. \onlinecite{sun2006examination}, Sun \emph{et al.}, using a model proposed by Guti{\'e}rrez, Taga and Johansson,\cite{gutierrez2001theoretical} have done a detailed study on vacancy distribution energies. Structures with all possible vacancy distributions within the chosen supercell were considered and compared. It was found that the supercell structure with two vacancies located at two octahedral sites about 8.2 \AA \/ away from each other, is the lowest in energy, and is in good agreement with experimental X-ray power diffraction data. Therefore we have chosen this structure as the defective spinel model for this study. A different study by Maglia and Gennari \cite{maglia2008energetics} produced results that are consistent with the calculations in Ref. \onlinecite{sun2006examination}.  The hydrogenated spinel structure of $\gamma$-alumina proposed by de Boer and Houben\cite{de1952proceedings} was confirmed by Sohlberg,\cite{sohlberg1999hydrogen} which corresponds to a stable and perfect hydrogenated spinel structure. The primitive cell contains eight O atoms, five Al atoms and one H atom.

The phonon dispersions were calculated by means of PHONOPY code,\cite{PhysRevB.78.134106} which is an implementation of post-process phonon analyzer, from the Hessian matrix calculated using density functional perturbation theory (DFPT) and PBE functional implemented in VASP. We used a 4$\times$1$\times$3, and a 2$\times$2$\times$2 supercell to calculate the eigenvalues of the Hessian matrix for the defective spinel structure and hydrogenated spinel structure, respectively. The phonon-related thermal properties of these compounds were then derived from the calculated phonon spectra.

\subsection{Evaluation of configurational entropy\label{II.b}}
The configurational (residual) entropy stems from the intrinsic disorder of the geometrically frustrated crystal lattice. A classical example is the case of H$_2$O ice that was first treated by Pauling.\cite{pauling1935structure} The hydrogen in ice has some degree of freedom sitting between the oxygen atoms, leading to an enormous number of configurations that are nearly indiscernible in energy. The presence of structural vacancies in a crystal also can increase the disorder and bring in considerable configurational entropy. 

The residual configurational entropies were evaluated using Pauling's approach.\cite{pauling1935structure} The procedure is described below for the two models of $\gamma$-alumina and for Boehmite. 
\\(a) Defective spinel structure of $\gamma$-alumina: Pauling's assumption that in the minimal primitive cell Al$_8$O$_{12}$ of $\gamma$-alumina structure the nine cation sites are occupied by eight Al$^{3+}$ ions completely at random gives the following estimate for the configurational entropy (per mole of Al$_2$O$_3$):
%%%%
\begin{equation} \label{eq:s0dm}
S^{\rm max}_{\rm cnf} = -\frac{9}{4} k_{\rm B} \left( \frac{1}{9} \ln \frac{1}{9} + \frac{8}{9} \ln \frac{8}{9} \right) \cong 0.785 k_{\rm B} ,
\end{equation}
where $k_{\rm B}$ is the Boltzmann constant. Since the present calculations show that cation vacancies prefer octahedral sites (six out of the nine sites per Al$_8$O$_{12}$ formula), the configurational entropy of $\gamma$-alumina at low temperatures reduces to:
%%%%
\begin{equation} \label{eq:s0do}
S^{\rm oct}_{\rm cnf} = -\frac{6}{4} k_{\rm B} \left( \frac{1}{6} \ln \frac{1}{6} + \frac{5}{6} \ln \frac{5}{6} \right) \cong 0.676 k_{\rm B} .
\end{equation}
%%%%
\\(b) Hydrogenated spinel structure of $\gamma$-alumina: A charge-balanced formula of the hydrogenated spinel, HAl$_5$O$_8$, corresponds to one protonated vacancy per six cation sites (two tetrahedral and four octahedral). The proton inside a cation vacancy binds to one of the neighboring O anions to form an OH group. Assuming a completely random substitution of any cation site by hydrogen (and also that the OH formation is completely uncorrelated), one arrives at the following upper-bound estimate $S^{\rm max}$ for the configurational entropy (per formula unit Al$_2$O$_3 \cdot 0.2$H$_2$O): 
%%%%
\begin{eqnarray} \label{eq:s0hm}
S^{\rm max}_{\rm cnf} & = & \frac{2}{5} k_{\rm B} \left[ \ln \frac{6^6}{5^5} + \left( \frac{2}{6} \ln4 + \frac{4}{6} \ln6 \right) \right] \\
& \cong & 1.744 k_{\rm B} , \nonumber
\end{eqnarray}
where the largest part, $1.081 k_{\rm B}$, is due to the randomness of vacancy distribution, while the remaining part, $0.663 k_{\rm B}$, is due to the proton disorder.
\\(c) Boehmite: The low-temperature phase of AlOOH has a layered structure in which hydroxylated double oxide layers are hold together by hydrogen bonds. The hydrogen bonds are arranged into zig-zag chains with two possible orientations of the hydroxyl groups, left $\cdots$H$-$O$\cdots$H$-$O$\cdots$ and right $\cdots$O$-$H$\cdots$O$-$H$\cdots$. The sequence can be reversed by means of a defect denoted here as a "break point" (BP). Let us denote the concentration of such point defects in Boehmite as $c=N_{\rm BP}/N = L^{-1}$. Here $L$ is the average length of  ordered (left or right) chains between two point defects, $N$ is the number of AlOOH units in the crystal, and $N_{\rm BP}$ is the number of BP defects. The defect formation energy $\Delta E$ is assumed to be positive. Free energy minimization with respect to the defect concentration  yields: $c=L^{-1} = \left[ 1+\exp(\Delta E/k_{\rm B}T) \right]^{-1}$. Thus, the equilibrium concentration of BP defects is exponentially small at low temperatures, which means that the residual entropy of Boehmite is zero. 
%%%%

\section{Results and Discussion\label{III}}
\subsection{Atomic structure}

\begin{figure}[htb]
\begin{center}
\includegraphics[width=8.7 cm]{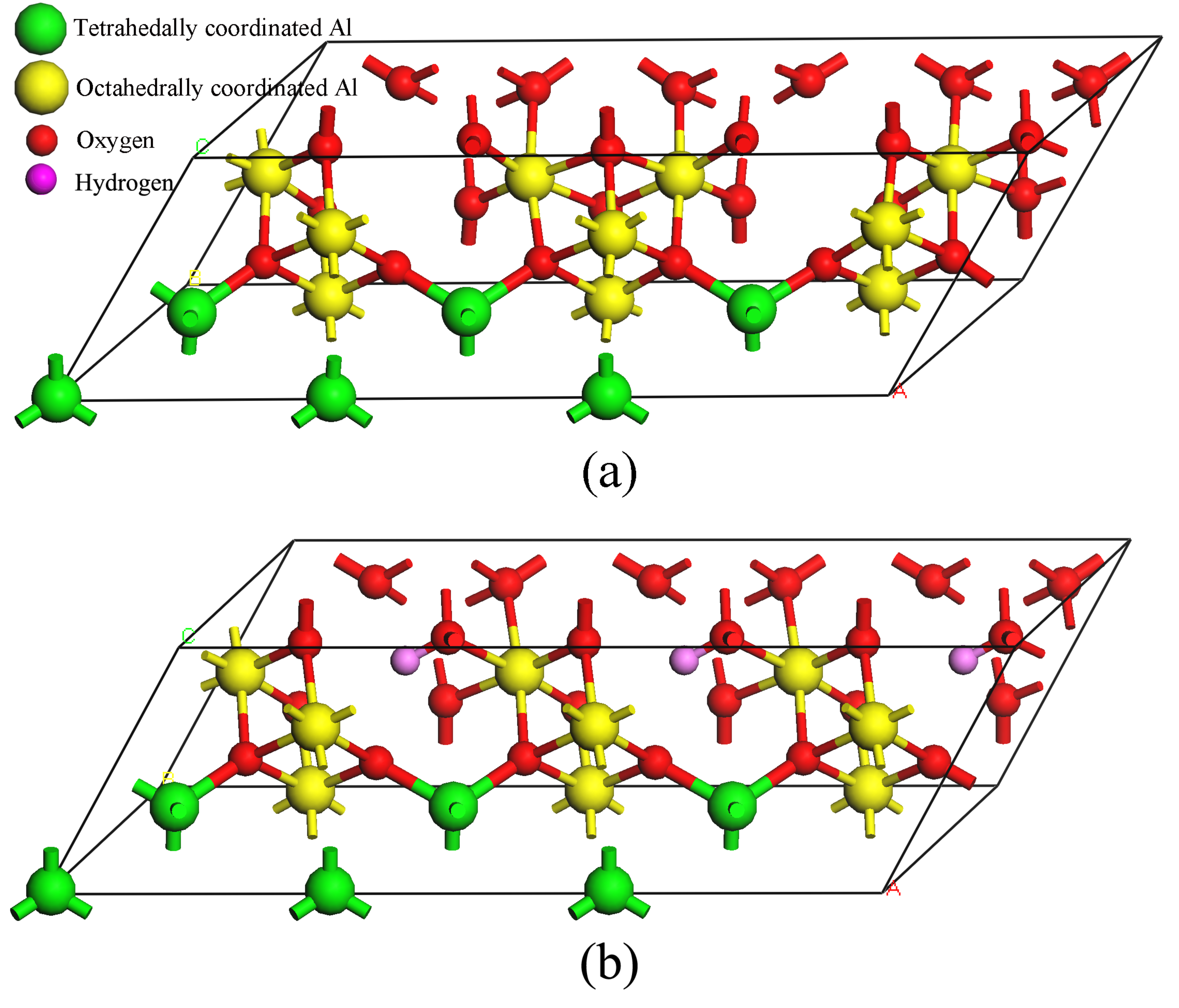}
\end{center}
\caption{\label{structure}Two structural models of $\gamma$-alumina: (a) the defective spinel model; (b) the hydrogenated spinel model with three primitive cells. For clarity, the Al atoms are shown larger than the O atoms.\label{structure}}
\end{figure}

Figure \ref{structure} shows the two structural models of $\gamma$-alumina. In the defective spinel model (Figure \ref{structure}(a)), there are 24 oxygen atoms that constitute the 6 tetrahedron cages and 12 octahedron cages, in which 16 aluminum atoms are located. Besides, there are two vacancies located inside two octahedron cages.  The primitive cell of a hydrogenated structure contains 14 atoms: 8 oxygen atoms, 5 aluminum atoms and 1 hydrogen atom. There are two choices for positioning the hydrogen atoms when constructing the structure of the hydrogenated model from the Mg$_2$Al$_4$O$_8$ primitive cell. The model can be built up either by replacing half of the Mg with Al and the other half with H, or by substituting one of the Al atoms with an H atom and also replacing all of the Mg by Al. In the first case, the hydrogen is tetrahedrally (TETRA) coordinated , while in the second case, the hydrogen is octahedrally (OCT) coordinated. However, during relaxation the hydrogen will deflect from the center of the oxygen cage to form a hydroxyl group with one of the neighboring oxygen atoms. We have found that the two structures (TETRA and OCT) are nearly isoenergetic (the calculated difference is less than 0.06 eV). Building a supercell by tripling the primitive cell of the hydrogenated model can slightly lower the total energy of the system due to the independent relaxation of the three hydrogens. However, the energy lowering is less than 0.1 eV. Figure \ref{structure}(b) shows the supercell of the hydrogenated spinel structure. The supercell was used to calculate the energy, but other properties were computed using a primitive cell. 

\begin{table}
 \caption{\label{v&b}Computed equilibrium volume $\Omega_0$ (per formula) and bandgap E$_g$ for defective and hydrogenated spinel $\gamma$-alumina and Boehmite.}
\begin{ruledtabular}
\begin{tabular} {l l l l}
Phase & Method & $\Omega_0$, \AA$^3$ & E$_g$, eV \\
\hline
\multirow{3}{*}{Al$_2$O$_3$} & PBE  & 47.373 & 3.91 \\
 & PBE0 & 45.885 & 5.96\\
 (Defective)& Exp. & 46.416$-$46.982\footnotemark[1]  & 2.5$-$8.7\footnotemark[2]\\
\hline
\multirow{3}{*}{Al$_2$O$_3 \cdot \frac{1}{5}$H$_2$O} & PBE  & 47.499 & 4.23\\
 & PBE0 & 45.991 & 6.35\\
 (Hydrogenated)& Exp. & 46.416$-$46.982\footnotemark[3]  & 2.5$-$8.7\footnotemark[4]\\
\hline
\multirow{3}{*}{AlOOH} & PBE  & 32.848 & 5.24\\
 & PBE0 & 31.560 & 7.59 \\
 (Boehmite)& Exp. & 32.650\footnotemark[5]& $-$\\
 \end{tabular}
\end{ruledtabular}
\footnotetext[1]{Refs. \onlinecite{zhou1991, krokidis2001theoretical, PhysRevB.70.125402}.}
\footnotetext[2]{Refs. \onlinecite{ealet1994electronic, kefi1993hybridization, snijders2002structure}.}
\footnotetext[3]{Refs. \onlinecite{zhou1991, krokidis2001theoretical, PhysRevB.70.125402}.}
\footnotetext[4]{Refs. \onlinecite{ealet1994electronic, kefi1993hybridization, snijders2002structure}.}
\footnotetext[5]{Ref. \onlinecite{christensen1982}.}
 \end{table}

The PBE-optimized structural data for the defective and hydrogenated spinel structures of $\gamma$-alumina are provided as Supplemental Material to this Article. The defective spinel model has a monoclinic symmetry group $C2/m$. The relaxed lattice parameters are in good agreement with previously published experimental and calculated data.\cite{zhou1991, krokidis2001theoretical, PhysRevB.70.125402} The calculated supercell equilibrium volume and bandgap of two structures of $\gamma$-alumina and Boehmite are given in Table \ref{v&b}. The difference between the calculated supercell equilibrium volume and experimental data is about 1\%.\cite{zhou1991, krokidis2001theoretical, PhysRevB.70.125402} The hydrogenated spinel structure has a triclinic symmetry group $P1$  due to the deviation of H atom from the cation vacancy center. The calculated supercell volume is also in good agreement with experimental data with about 1\% overestimation.\cite{zhou1991, krokidis2001theoretical} The equilibrium volumes calculated with PBE0 are also very close to the experimental data.

The $\gamma$-alumina is an insulator with a wide bandgap, of about 6.8 eV, which is smaller than the bandgap of $\alpha$-alumina.\cite{kefi1993hybridization}  Due to the complexity of its bulk and surface structures, the bandgap of $\gamma$-alumina poses a challenge for its experimental determination.\cite{ealet1994electronic, kefi1993hybridization, snijders2002structure} Thus, the experimentally obtained bandgap values vary considerably from 2.5 to 8.7 eV. The bandgap of $\gamma$-alumina and Boehmite calculated in this work are also summarized in Table I. The PBE0 gives higher and more accurate bangaps for these three structures. 

\begin{figure}[htb]
\begin{center}
\includegraphics[width=8.7 cm]{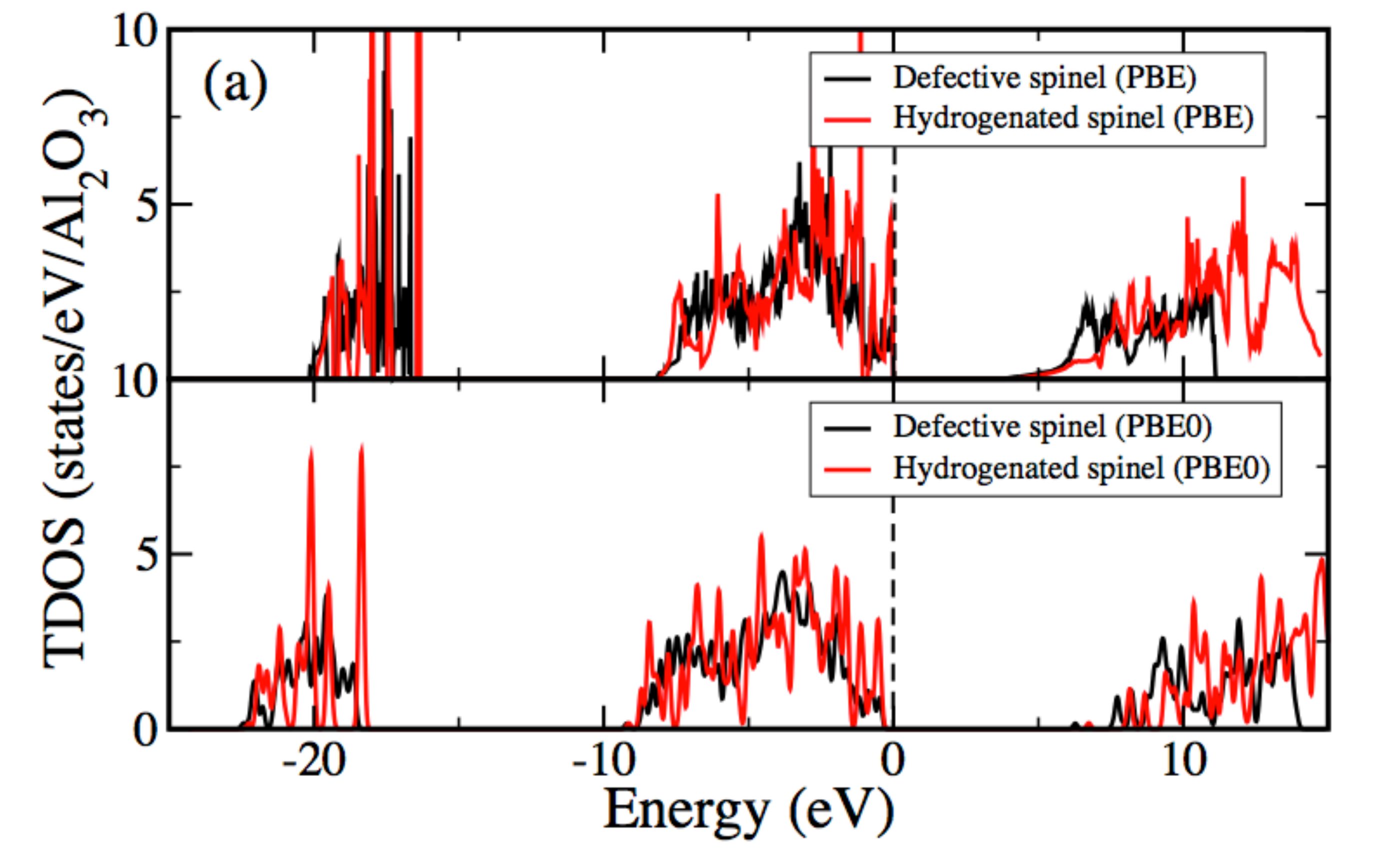}
\includegraphics[width=8.7 cm]{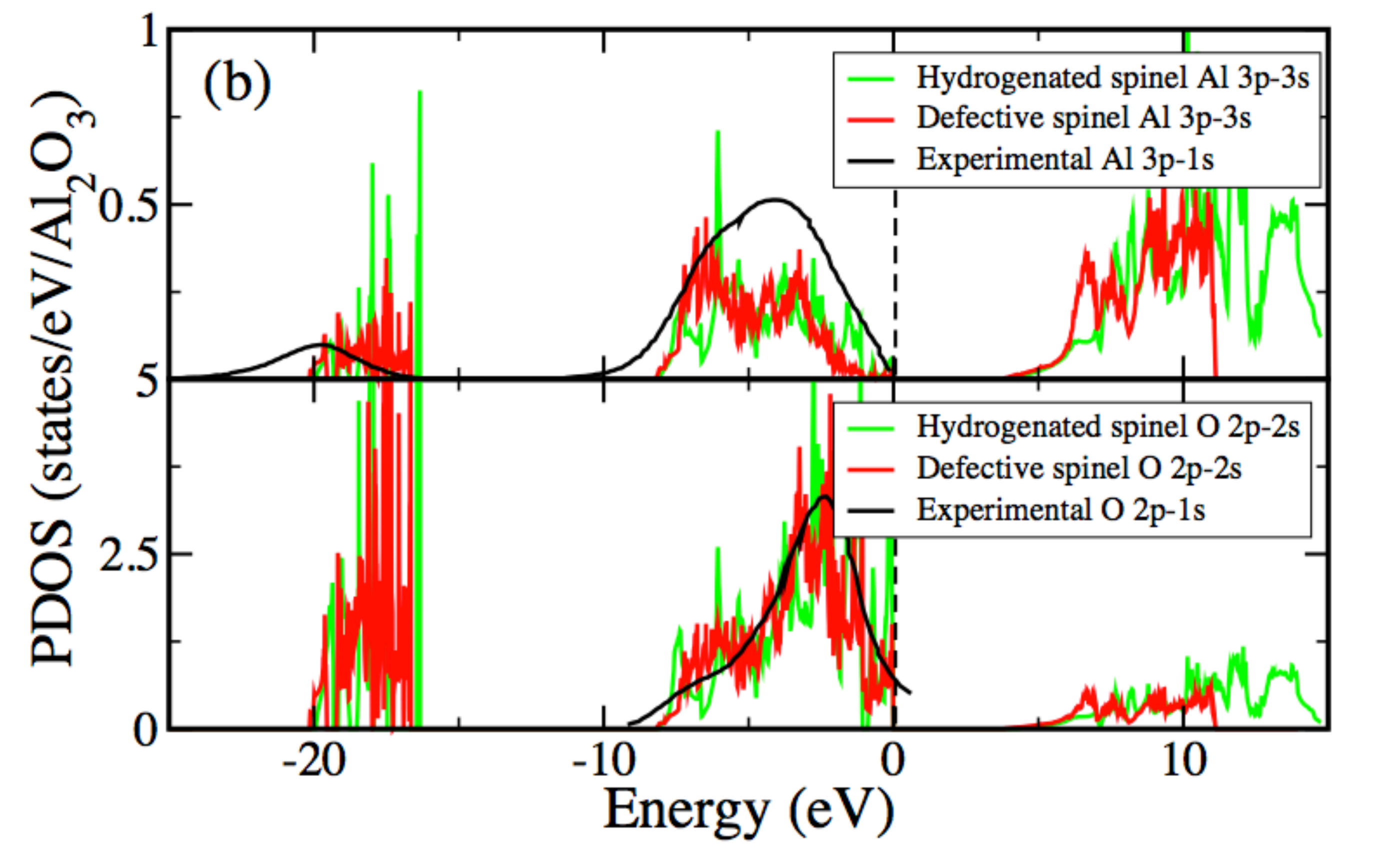}
\end{center}
\caption{ \label{dos}Density of electron states of $\gamma$-alumina: (a) PBE-calculated and PBE0-calculated TDOS; (b) PBE-calculated PDOS compared with data from X-ray emission spectroscopy experiments in Ref. \onlinecite{kefi1993hybridization}.} 
\end{figure}

As can be seen in Figure \ref{dos}(a), the two used functionals present similar total density of states (TDOS) for $\gamma$-alumina. The peaks are almost coincident within the whole energy range for the two structural models. As expected, the valence band region is separated into two parts: the upper valence band (UVB) ranging from the Fermi level down to $-9$ eV below this, and the lower valence band (LVB) in the energy range from $-16$ eV to $-21$ eV. The calculated TDOS is in good agreement with previously published experimental and calculated data.\cite{ealet1994electronic, kefi1993hybridization, PhysRevB.70.125402, PhysRevB.78.014106} Figure \ref{dos} (b) shows the projected density of states (PDOS) (calculated using PBE) in comparison with experimental data. The corresponding densities of states are projected from the Kohn-Sham wave functions onto atomic Bader volumes and calculated within these volumes. The calculated UVB of PDOS matches very well with the data retrieved from Ref. \onlinecite{kefi1993hybridization}, but the calculated LVB deviates slightly from the experimental data. The LVB obtained with PBE0 matches better with the experimental data than that obtained with PBE. The UVB is dominated by the 2$p$ orbitals of O and has small contributions from Al 3$s$ and Al 3$p$ orbitals, which has also been confirmed experimentally by Ealet \emph{et al.}\cite{ealet1994electronic} In UVB, the calculated data has two peaks centered at $-3$ eV and $-7$ eV which reflect the overall contributions of orbitals of Al and O. The two-peak structure is not visible in the experimental data of Ref. \onlinecite{kefi1993hybridization}. However, it is stated in Ref. \onlinecite{ealet1994electronic} that these two peaks in UVB are separated and correspond to the O 2$p$ bonding orbital ($-$7 eV) and O 2$p$ non-bonding orbital ($-$3 eV). Therefore, it can be seen that there is extensive hybridization between the O 2$p$ and Al 3$p$ orbitals. This is a result of the covalent bonding between Al and O in $\gamma$-alumina. In LVB, the main contribution comes from the O 2$s$ orbitals. This has been also verified in Ref. \onlinecite{ealet1994electronic}.

\begin{figure}[htb]
\begin{center}
\includegraphics[width=8.7 cm]{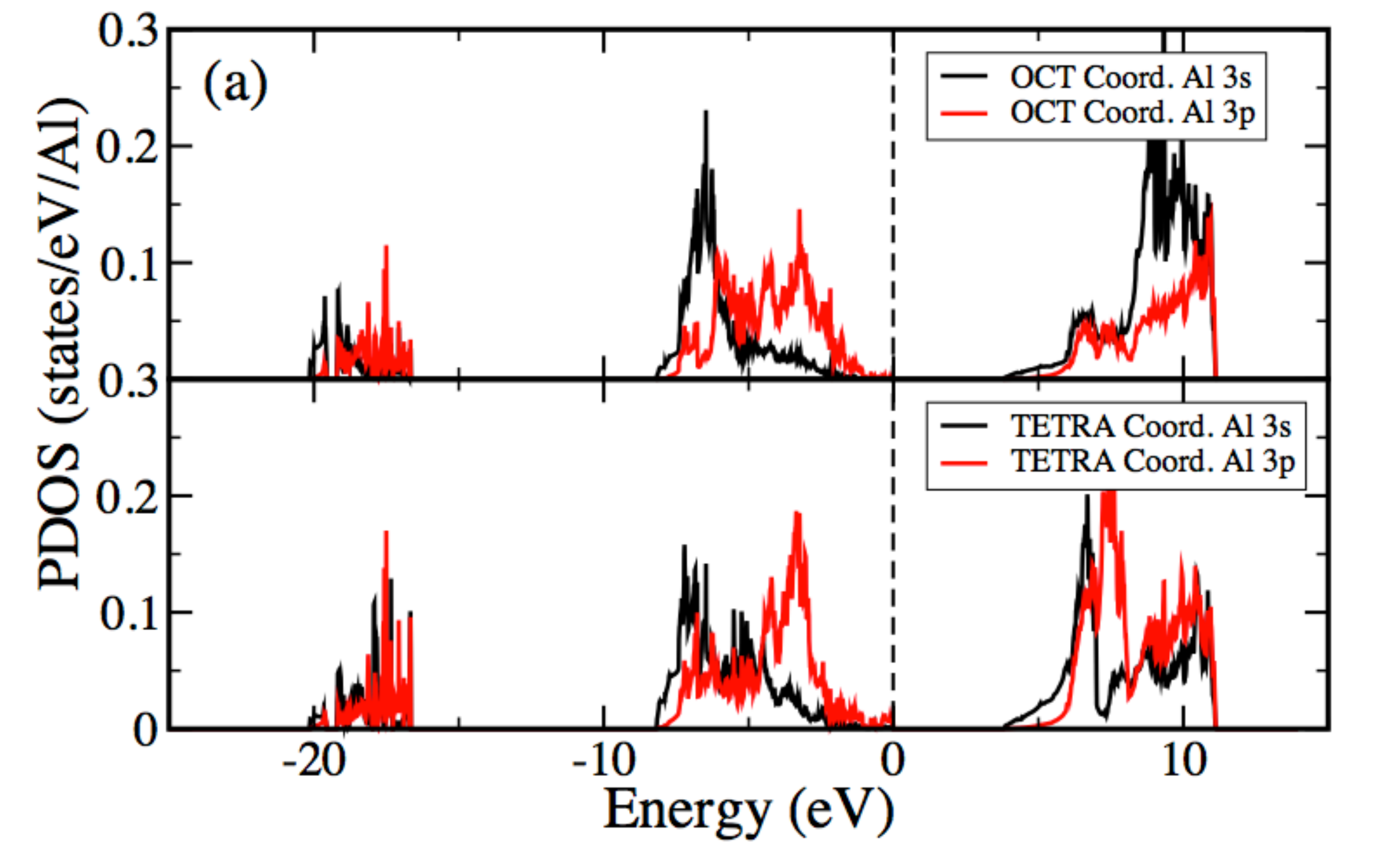}
\includegraphics[width=8.7 cm]{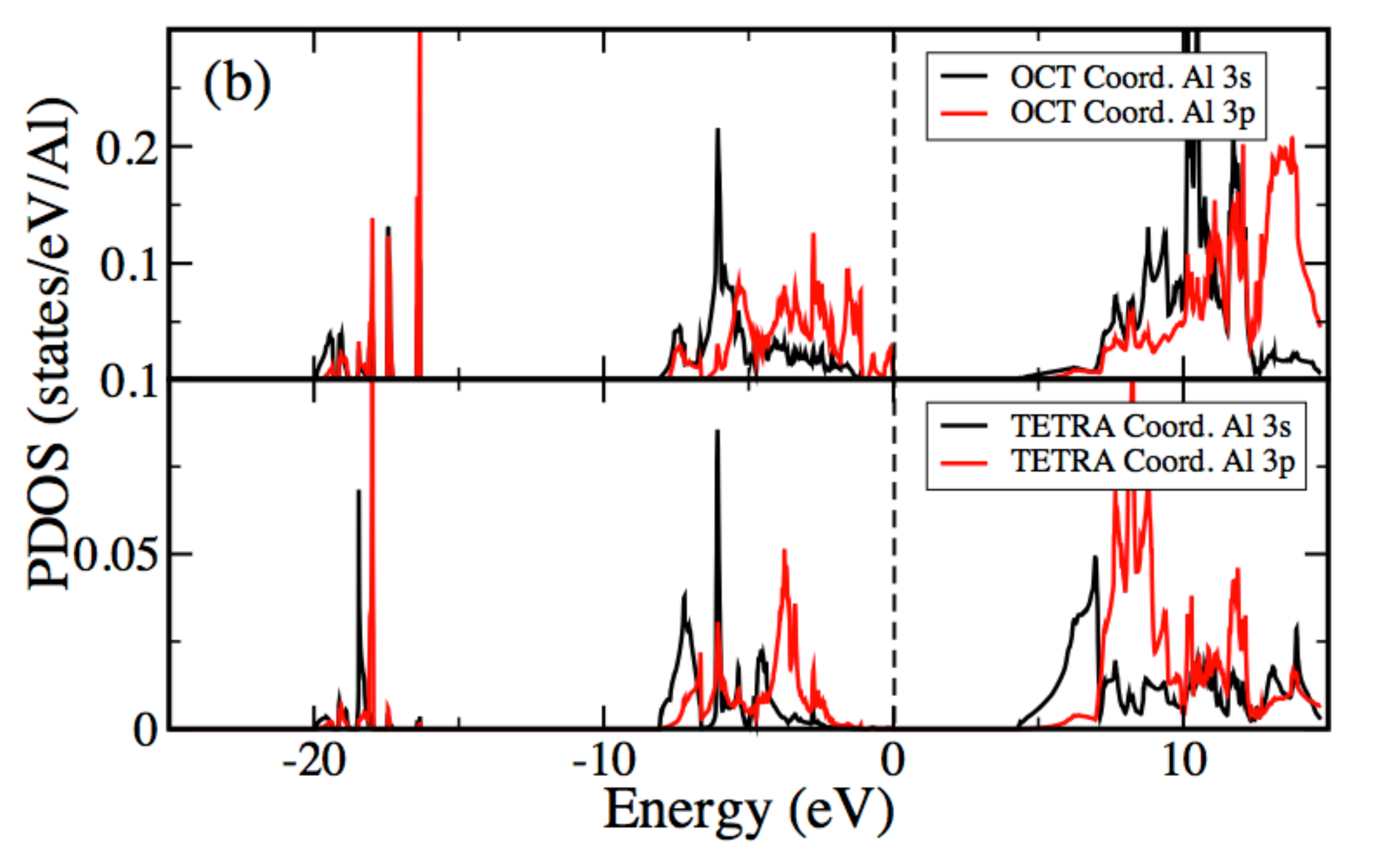}
\end{center}
\caption{\label{pdos}PBE-calculated PDOS on Al in $\gamma$-alumina for: (a) the defective spinel model; (b) the hydrogenated spinel model.}
\end{figure}

As one can see from Figures \ref{pdos} (a) and (b), the Al 3$s$ and Al 3$p$ orbitals are extensively hybridized due to the polycoordination. The Al 3$s$ and 3$p$ orbitals hybridize over the whole valence band and beyond. By comparing Figure \ref{pdos} (a) and  Figure \ref{pdos} (b), it is visible that such hybridization is more extensive in the defective spinel phase than in the hydrogenated spinel phase. Another noticeable feature in the data shown in Figure \ref{pdos} is that the hybridization is more extensive on tetrahedrally coordinated Al atoms than in octahedrally coordinated Al.

\subsection{Dynamic stability and phonon spectra}
The dynamic stabilities of previously proposed structures of $\gamma$-alumina have been investigated only at the $\Gamma$ point,\cite{sohlberg1999hydrogen, wolverton2000phase, sun2006examination} 
%%Thus, the dynamic stability of this material has not been confirmed by full-spectrum phonon dispersion calculations. 
To the best of our knowledge, only the stability of the defective spinel phase proposed by Guti{\'e}rrez, Taga and Johansson\cite{gutierrez2001theoretical} has been confirmed by full-spectrum phonon dispersion calculations in Ref. \onlinecite{PhysRevB.78.014106}. In Ref. \onlinecite{PhysRevB.78.014106}, Ching and coworkers used ultrasoft pseudopotentials and a finite-displacement method for phonon calculations. Loyola have employed classic force field methods to calculate the phonon DOS of several models including the defective spinel phase in Ref. \onlinecite{loyola2010atomistic}, and compared the result with that of Ref. \onlinecite{PhysRevB.78.014106}. Some differences between the classical and \emph{ab initio} calculated DOS have been found for frequencies higher than 200 cm$^{-1}$. It has been suggested that the \emph{ab initio} results are less accurate due to the limitations imposed by the usage of the finite-displacement method.\cite{loyola2010atomistic} 

\begin{figure}[htb]
\begin{center}
\includegraphics[width=8.7 cm]{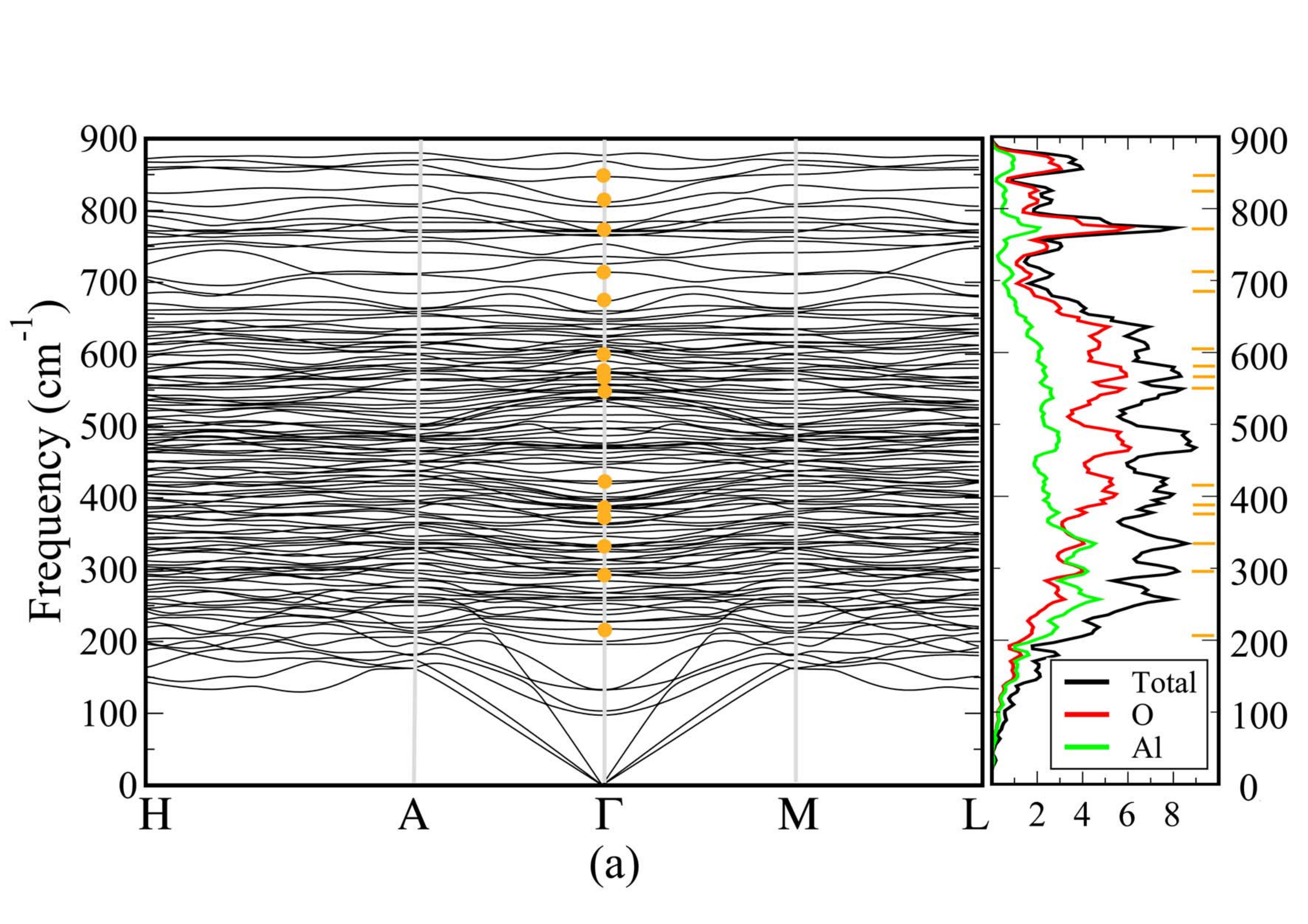}
\includegraphics[width=8.7 cm]{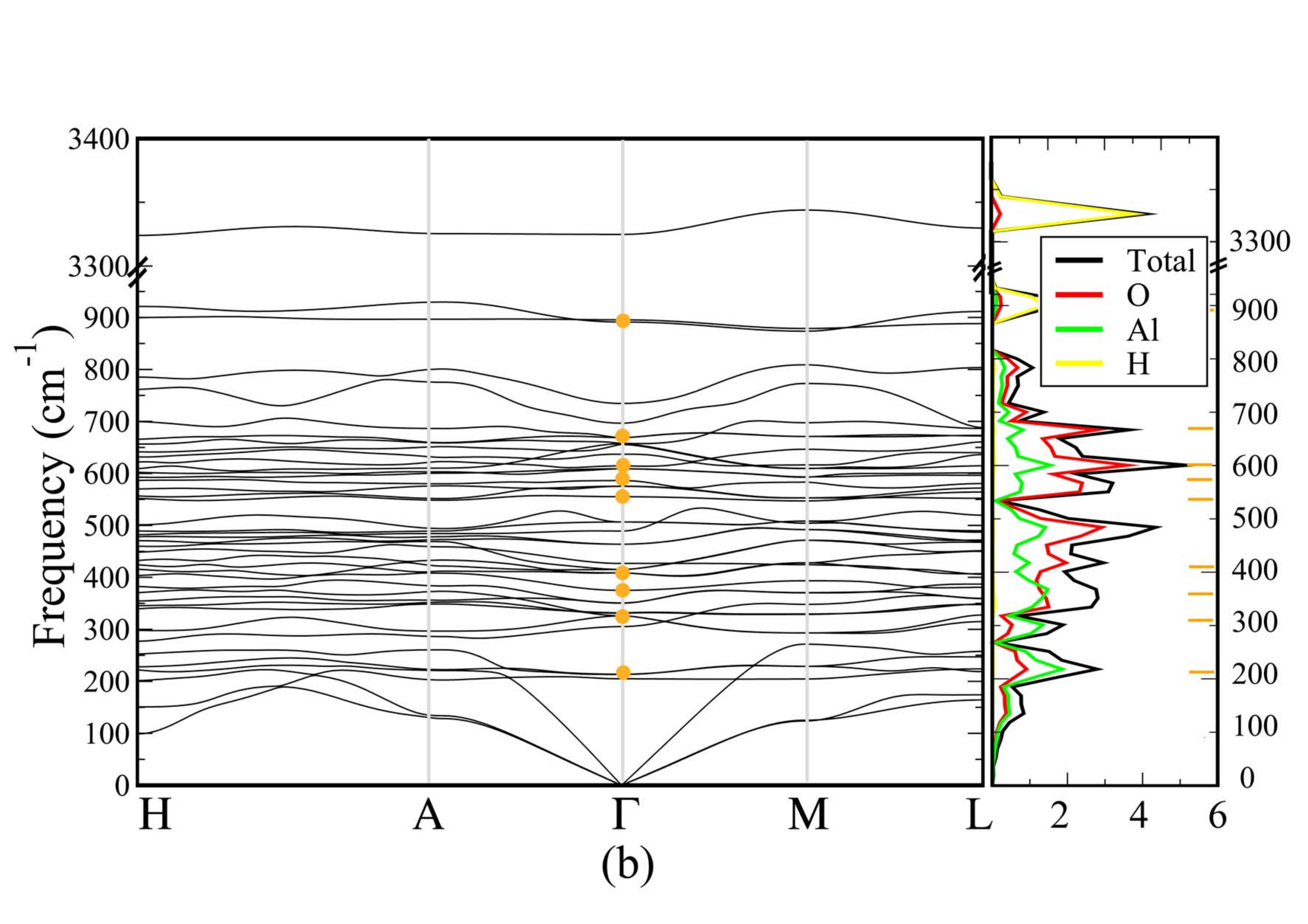}
\includegraphics[width=8.7 cm]{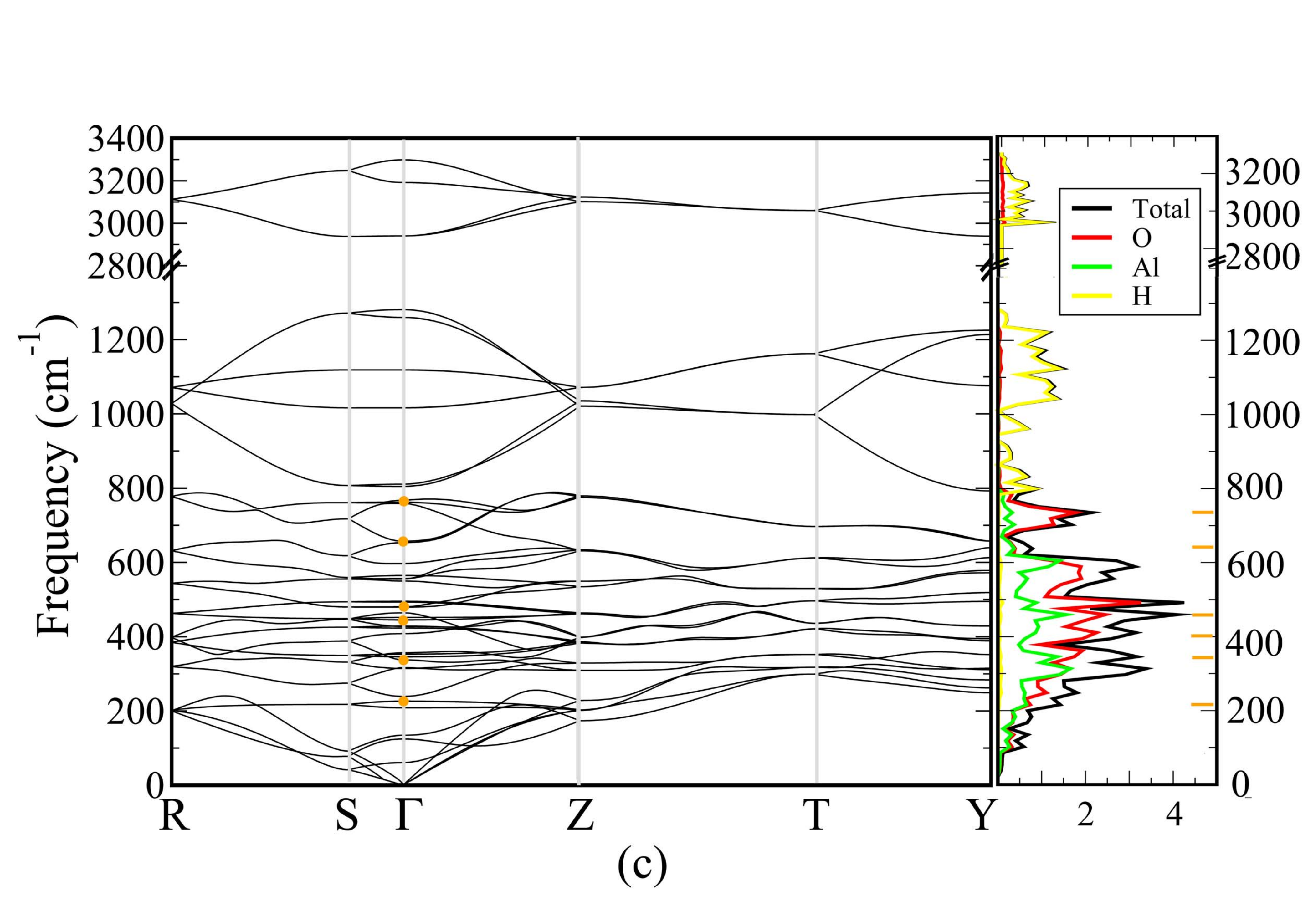}
\end{center}
\caption{\label{phonon}Phonon dispersion and phonon DOS of (a) the defective spinel $\gamma$-alumina; (b) the hydrogenated spinel $\gamma$-alumina; (c) Boehmite AlOOH. The orange dots and marks in are indicating calculated Raman-active modes and peaks.}
\end{figure}

We use PAW potentials and the DFPT method, as described in the methodology section, to calculate the phonon spectra of the defective spinel model and of the hydrogenated model. The obtained phonon dispersion curves and the phonon DOS are shown in Figure \ref{phonon}. The phonon dispersion curves display no imaginary frequencies, suggesting the dynamic stability of the considered structures. For the structures of $\gamma$-alumina, the acoustic modes range from 0 to 200 cm$^{-1}$, and have equivalent contributions of  Al and O in both models. The optical branches are rather flat, and the O atoms contribute more than Al for frequencies above 350 cm$^{-1}$. The O atom is lighter in mass than Al, so it makes a larger contribution in the higher frequency range. The most prominent peak at 500 cm$^{-1}$ is attributed to the stretching mode of the Al-O bond in the octahedral AlO$_6$ unit. The noticeable differences between the two models are well evident for frequencies over 700 cm$^{-1}$. Between 700 and 1000 cm$^{-1}$, the defective model gives rise to 5 peaks, while only four peaks are visible for the hydrogenated model. Besides, only two peaks that are located at 750 and 770 cm$^{-1}$ coincide with each other. In both models, these two peaks can be attributed to the Al-O bonds in the tetrahedral AlO$_4$ unit. The peaks of the hydrogenated model at 900 and 3330 cm$^{-1}$ are attributed to the vibrational modes of the hydroxyl groups, where the proton contribution is more significant than that of O atoms. 

The calculated phonon DOS for the defective spinel is in good agreement with the data of Ref. \onlinecite{PhysRevB.78.014106}. The phonon DOS extinguishes at 880 cm$^{-1}$ in the \emph{ab initio} calculations of Ching and coworkers.\cite{PhysRevB.78.014106} According to our results, the phonon DOS is also terminated at the same frequency. We note that, in Ref.65, the phonon DOS gradually extinguishes and eventually vanishes at 980 cm$^{-1}$.

The calculated Raman-active modes are marked by filled circles and streaks in Figure \ref{phonon}. Due to the poor crystallinity, Raman spectrum of $\gamma$-alumina with clear features is difficult to obtain experimentally.\cite{roy1995phonons, ruan2001comparison, roy1995phonons, aminzadeh1999raman, otto1992raman, dyer1993surface} In contrast, experimental Raman spectra of Boehmite are quite clear.\cite{ruan2001comparison, roy1995phonons} For this material, there are three peaks at 360 cm$^{-1}$,  497 cm$^{-1}$  and 677 cm$^{-1}$, as reported by Roy and Sood.\cite{roy1995phonons}  Three additional peaks at 228 cm$^{-1}$, 451 cm$^{-1}$ and 732 cm$^{-1}$ were observed by Ruan \emph{et al.} The frequencies of Raman-active modes obtained in our calculations are at 220 cm$^{-1}$, 357 cm$^{-1}$, 460 cm$^{-1}$, 497 cm$^{-1}$, 680 cm$^{-1}$ and 730 cm$^{-1}$. These values agree well with experimental Raman peak frequencies.\cite{ruan2001comparison} As regard to the $\gamma$-alumina, the defective spinel phase shows more features than the hydrogenated phase. The Raman-active modes calculated for the hydrogenated spinel phase can find their correspondence in the defective spinel phase.

\subsection{Thermal stability}

In this section we analyse the thermal stabilities of the two structural models (defective and hydrogenated spinel) of $\gamma$-alumina, relative to water Boehmite, on the basis of Gibbs free energy evaluation. However, since the energies of compounds with different stoichiometries cannot be compared directly, we compare the Gibbs free energy of Al$_2$O$_3$+$n$H$_2$O (a two-phase mixture of defective spinel $\gamma$-alumina and H$_2$O) with the hydrogenated spinel phase HAl$_5$O$_8$ and Boehmite AlOOH, as it was previously done by Wolverton and Hass\cite{wolverton2000phase} and also by Sun \emph{et al.},\cite{sun2006examination} using the following reactions:\\
\begin{equation} \label{chm:b2da}
\mathrm {2AlOOH \longrightarrow Al_2O_3 + H_2O}
\end{equation}
and \\
\begin{equation} \label{chm:b2ha}
\mathrm{5AlOOH \longrightarrow HAl_5O_8 + 2H_2O} .
\end{equation}
The reactions (\ref{chm:b2da}) and (\ref{chm:b2ha}) correspond to the decomposition of Boehmite into defective spinel phase plus H$_2$O and hydrogenated phase plus H$_2$O, respectively. Therefore, the Gibbs free energy changes ($\Delta G$) for the reactions (\ref{chm:b2da}) and (\ref{chm:b2ha}) can be expressed as
\begin{equation} \label{eq:g1}
\Delta G_1 = G(\mathrm{Al_2O_3}) + G(\mathrm{H_2O}) - 2G(\mathrm{AlOOH})
\end{equation}
and\\
\begin{equation} \label{eq:g2}
\Delta G_2 = G(\mathrm{HAl_5O_8})+2G(\mathrm{H_2O})-5G(\mathrm{AlOOH}) 
\end{equation}
where $\Delta G_1$ and $\Delta G_2$ refer to the free energy change for the processes in reaction (\ref{chm:b2da}) and in reaction (\ref{chm:b2ha}) respectively, and $G(\mathrm{Al_2O_3})$, $G(\mathrm{HAl_5O_8})$, $G(\mathrm{AlOOH})$, and $G(\mathrm{H_2O})$ are the Gibbs free energies of the corresponding materials at the temperature of interest.

Besides, the relative stabilities of the defective and hydrogenated spinel phases can be evaluated without involving Boehmite, via the reaction
\begin{equation} \label{chm:ha2da}
\mathrm {2HAl_5O_8 \longrightarrow 5Al_2O_3 + H_2O} .
\end{equation}
The corresponding Gibbs free energy change is then:
\begin{equation} \label{eq:g3}
\Delta G_3 = G(\mathrm{Al_2O_3}) + 1/5 G(\mathrm{H_2O}) - 2/5 G(\mathrm{HAl_5O_8})
\end{equation}
The expression for the Gibbs free energy is\\
\begin{equation} \label{eq:gpt}
G(P,T) = E + P V - T S 
\end{equation}
Here $E$ is the internal (total) energy, $P$ is the pressure, and $V$ is the volume. Under the Born-Oppenheimer approximation, the internal energy decomposes into two parts: the electronic energy $E_\mathrm{{el}}(V,T)$ and vibrational energy $E\mathrm{_{vib}}(V,T)$. As the compounds of our interest here are insulators, the electronic entropy contribution can be neglected in the considered temperature range. For the considered compounds, the entropy also contains another important term, the configurational entropy $S_\mathrm{{cnf}}$. For condensed phases at low pressure, or for systems without noticeable $P V$ change, the term $P V$ can be neglected. Then the Gibbs free energy expression for the considered compounds of aluminum becomes
\begin{equation} \label{eq:gfin}
G(0,T) = E\mathrm{_{el}}(V_0) + E\mathrm{_{vib}}{(V_0,T)} - T \cdot (S\mathrm{_{vib}}+S\mathrm{_{cnf}})
\end{equation}
where the ground-state total electronic energy $E\mathrm{_{el}}(V_0)$ is calculated for the optimized volume $V_0$ using VASP within the framework of DFT using the PBE or PBE0 functionals. The vibrational energy $E_\mathrm{{vib}}(V_0,T)$ and vibrational entropy $S\mathrm{_{vib}}$ were calculated using the PHONONPY code within the quasi-harmonic approximation (QHA). The Gibbs free energy of H$_2$O, in the gas and liquid state, was evaluated starting from the PBE calculated electronic and zero-point energy of an H$_2$O molecule, and adding temperature-dependent contributions (including the translational, vibrational and rotational free energy, vaporization energy, and the $P V$ term) retrieved from the NIST online database.\cite{mallard1992nist}  The entropy of solid-state H$_2$O was extrapolated based on the data at 0 K and the data above 273.15 K by means of polynomial fitting. 

\begin{figure}[h1tb]
\begin{center}
\includegraphics[width=8 cm]{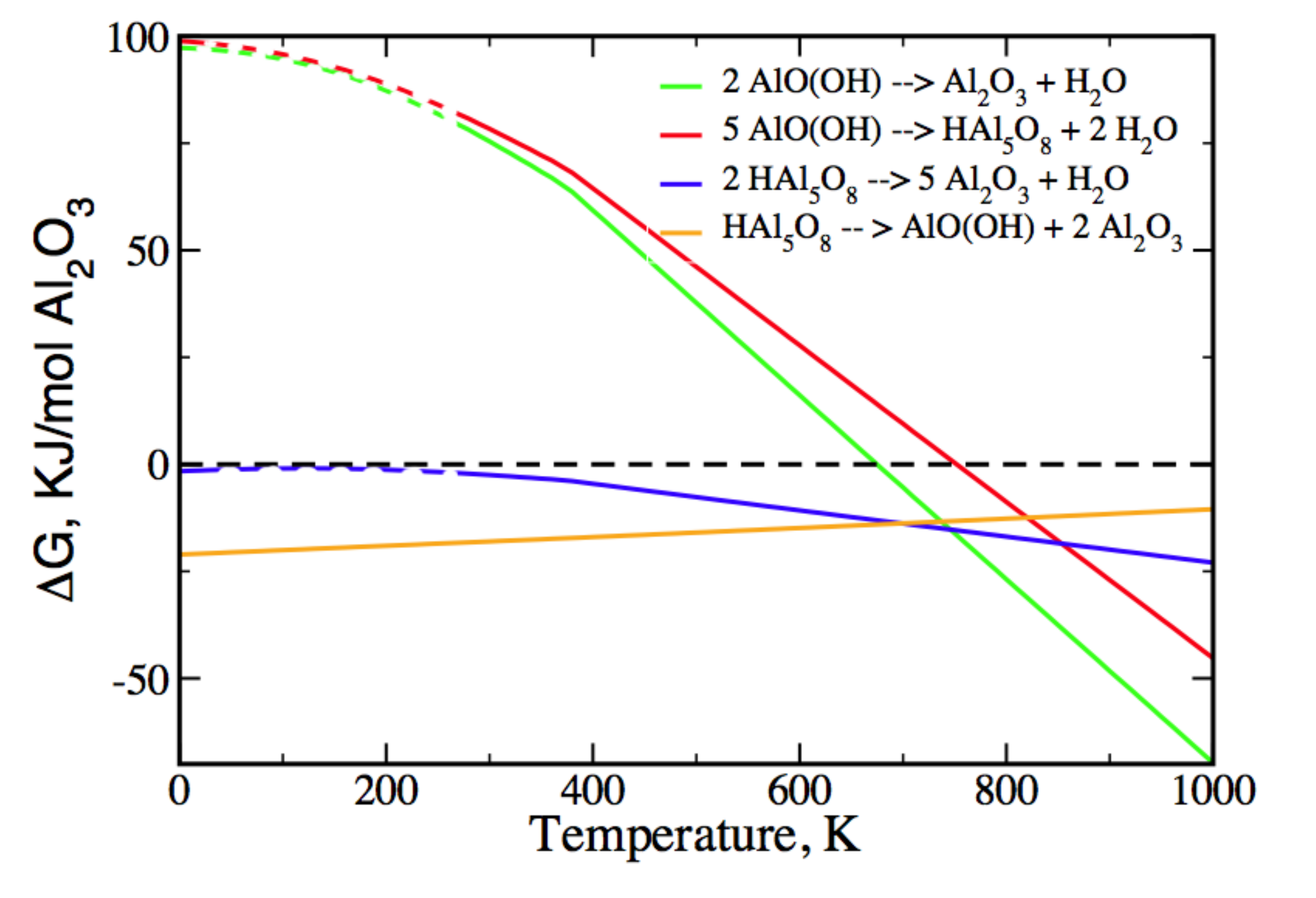}
\end{center}
\caption{\label{gibbs}Calculated Gibbs free energy changes: $\Delta G_1$ (green line) for the decomposition of Boehmite into defective $\gamma$-alumina and H$_2$O; $\Delta G_2$ (red line) for the decomposition of Boehmite into hydrogenated $\gamma$-alumina and H$_2$O; $\Delta G_3$ (blue line) for the decomposition of hydrogenated $\gamma$-alumina into defective $\gamma$-alumina and H$_2$O; and $\Delta G_4$ (yellow line) for the formation of Boehmite and defective $\gamma$-alumina from hydrogenated $\gamma$-alumina. The dotted parts of lines show extrapolated data.}
\end{figure}

The calculated temperature dependencies of the Gibbs free energy changes for reactions (\ref{eq:g1}), (\ref{eq:g2}) and (\ref{eq:g3}) are shown in Figure \ref{gibbs}. Considering the change in the Gibbs free energy for the reaction (\ref{eq:g1}) of decomposition of Boehmite into defective $\gamma$-alumina and water, the calculated critical temperature for the reaction is 678 K, in good agreement with the reaction temperature range from 673.15 to 723.15 K determined experimentally.\cite{alteo} Previous theoretical values systematically overestimated this critical temperature by 100 K\cite{krokidis2001theoretical, sun2006examination} The underlying reason is that the previous calculations did not take into account the vibrational and configurational contributions to the free energy. Neglecting these two terms will increase the Gibbs free energy change of reaction (\ref{eq:g1}) and increase the value of the critical temperature. The good agreement with experimental data indicates good accuracy of the methodology followed in this work.

The improved accuracy enables for a detailed analysis of the relative thermal stabilities of the $\gamma$-alumina models. The Gibbs free energy change of reaction (\ref{eq:g2}) is positive up to $T = 753$ K, as indicated by the red curve in Figure \ref{gibbs}. It suggests that the spontaneous decomposition of Boehmite into the hydrogenated spinel phase should not occur below 753 K. However, the transformation of the Boehmite phase into the defective spinel phase becomes thermodynamically favorable already at temperatures above 678 K. This result shows that the defective spinel phase, rather than the hydrogenated spinel phase, is the ground state of $\gamma$-alumina. Furthermore, the Gibbs free energy of the reaction (\ref{eq:g3}) never becomes positive in the whole temperature range of interest, and beyond. It suggests the instability of the hydrogenated spinel phase in comparison with the defective spinel phase. As such, the hydrogenated-spinel alumina phase will spontaneously decompose into the defective spinel phase and water. We also used the energies from PBE0 type of calculations at 0 K to assess the Gibbs free energy in reactions (\ref{eq:g1}), (\ref{eq:g2}), and (\ref{eq:g3}). The result is consistent with the PBE type of calculations. Our result is in agreement with that of Ref. \onlinecite{wolverton2000phase}, and disagrees with the conclusion of Ref. \onlinecite{sun2006examination} about the stability of hydrogenated spinel phase. 

The above analyses involved the experimental thermochemical data for H$_2$O, and, therefore, were not fully \textit{ab initio}. The assessment of relative stabilities of these materials without involving H$_2$O can be made. The composition of hydrogenated spinel $\gamma$-alumina can be expressed as the mixture of Boehmite and defective spinel $\gamma$-alumina
\begin{equation} \label{chm:ha2dab}
\mathrm {HAl_5O_8 \longrightarrow 2 Al_2O_3 + AlOOH} .
\end{equation}
Then the Gibbs free energy of hydrogenated spinel $\gamma$-alumina can be compared with that of the mixture of Boehmite and defective spinel $\gamma$-alumina,\\
\begin{equation} \label{eq:g4}
\Delta G_4 = 2 G(\mathrm{Al_2O_3}) + G(\mathrm{AlOOH}) - G(\mathrm{HAl_5O_8}) .
\end{equation}
This excludes the strongly temperature dependent contribution from H$_2$O. The result is shown in Figure \ref{gibbs}. The Gibbs free energy of hydrogenated $\gamma$-alumina is always higher than that of the mixture of Boehmite and defective $\gamma$-alumina, which confirms the instability of the hydrogenated $\gamma$-alumina in comparison with the defective spinel phase.

Worth of notice is that, we stick to the lower bound of the residual configurational entropy of Boehmite (zero at low temperatures). However, the defect formation energy $\Delta E$ introduced at the end of Sec. \ref{II.b} (which can be estimated by comparing the energies of proton-ordered Boehmite and proton-disordered Diaspore) in Boehmite is small, which means additional configurational entropy of AlOOH at high temperatures. In that sense, the Gibbs free energy change $\Delta G_4$ would be more negative and the hydrogenated spinel phase more unstable. The accurate description of the entropy gain with temperature increase and subsequent effects on the transformation of Boehmite will be reported in a separate study, since the current work focuses on $\gamma$-alumina. 

\begin{figure}[h1tb]
\begin{center}
\includegraphics[width=8.5 cm]{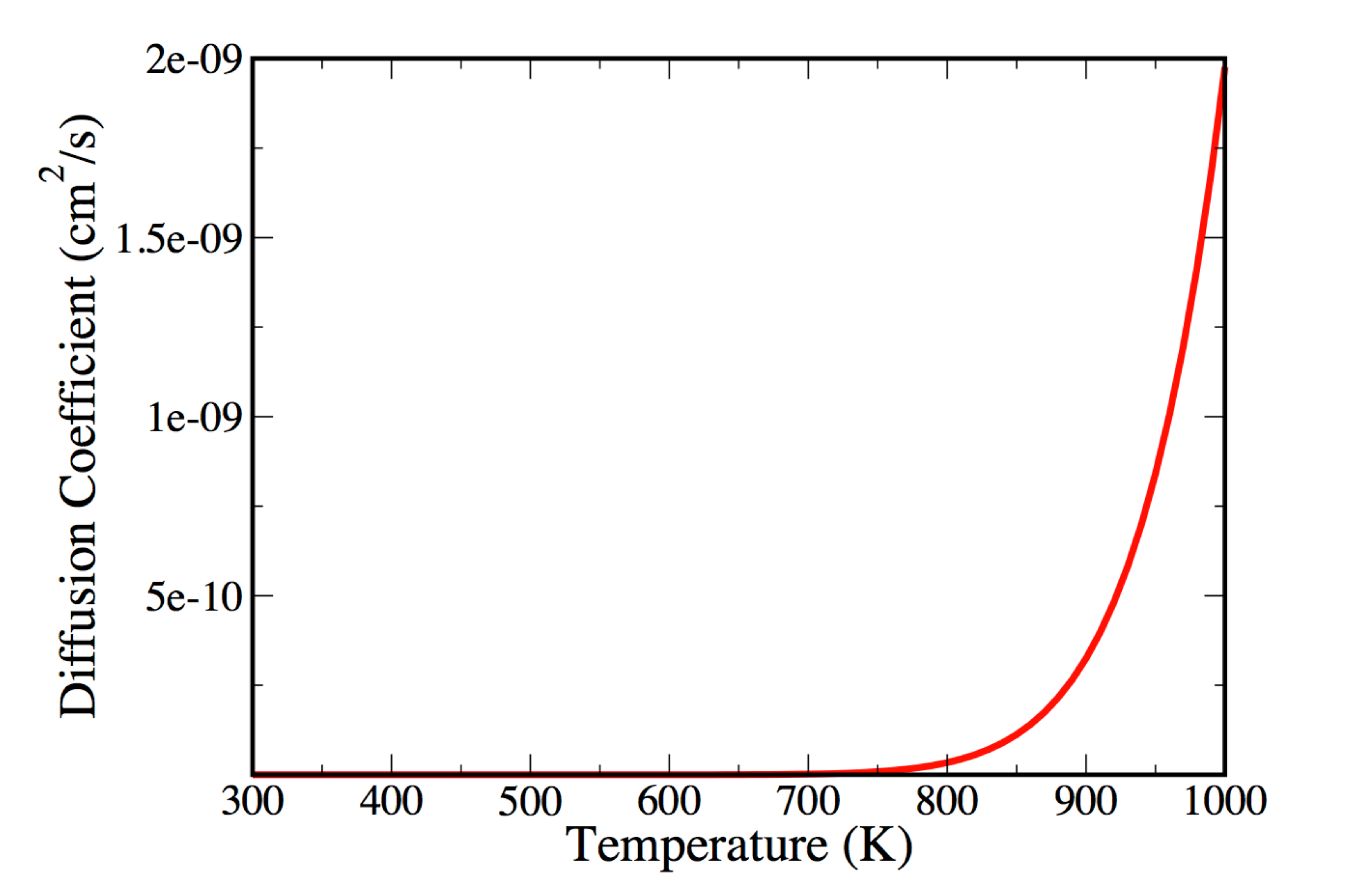}
\end{center}
\caption{\label{h-hoping}Hydrogen diffusion coefficient in the hydrogenated spinel phase of $\gamma$-alumina.}
\end{figure}

Nevertheless, it is likely that the hydrogenated spinel phase will not fully transform into the defective spinel phase above 753 K, since the vacancies are good traps for H and there is always bulk hydrogen detected in samples. Sohlberg \emph{et al.}\cite{sohlberg1999hydrogen} calculated the potential energy profile for hydrogen in $\gamma$-alumina to obtain a migration battier of about 1.4 eV. The diffusion coefficient of H in the hydrogenated $\gamma$-alumina can be estimated using a classical expression
\begin{equation} \label{eq:dh}
D = \frac{\rho^2\cdot\nu}{2k_{\rm B}T} \int_{E_m}^{\infty} e^{-\varepsilon/k_{\rm B}T}d\varepsilon = \frac{\rho^2\cdot\nu}{2} \cdot e^{-E_m/k_{\rm B}T}
\end{equation}
where $\nu$ (3330 cm$^{-1}$)is the frequency of stretching motion of hydrogen in a hydroxyl group, $\rho$ (1.5 \AA) is the hopping distance, and $E_m$ is the migration barrier. Figure \ref{h-hoping} shows the estimated diffusion coefficient of H in the hydrogenated spinel phase. It should be noted that the diffusion of hydrogen only becomes significant at temperatures above 900 K. For lower temperatures, the diffusion rate of H is extremely slow. That may be the main reason why the hydrogenated spinel phase is seen more often than the defective spinel phase in laboratory environments. We therefore conclude that the bulk H detected in experimental investigations of $\gamma$-alumina is not intrinsic in nature. At the same time, the surface H can still be natural for $\gamma$-alumina. As Digne \emph{et al.} have shown, the surface adsorption of hydrogen is energetically favorable and the surface hydroxyl groups remain stable even at high temperature.\cite{digne2002hydroxyl}  In $\theta$-alumina, which bears great similarity with $\gamma$-alumina, it is found that some facets have negative surface energy and stay hydroxylated up to high temperatures.\cite{topsoe2004negative}

\subsection{Thermal properties}

\begin{figure}[h1tb]
\begin{center}
\includegraphics[width=8.7 cm]{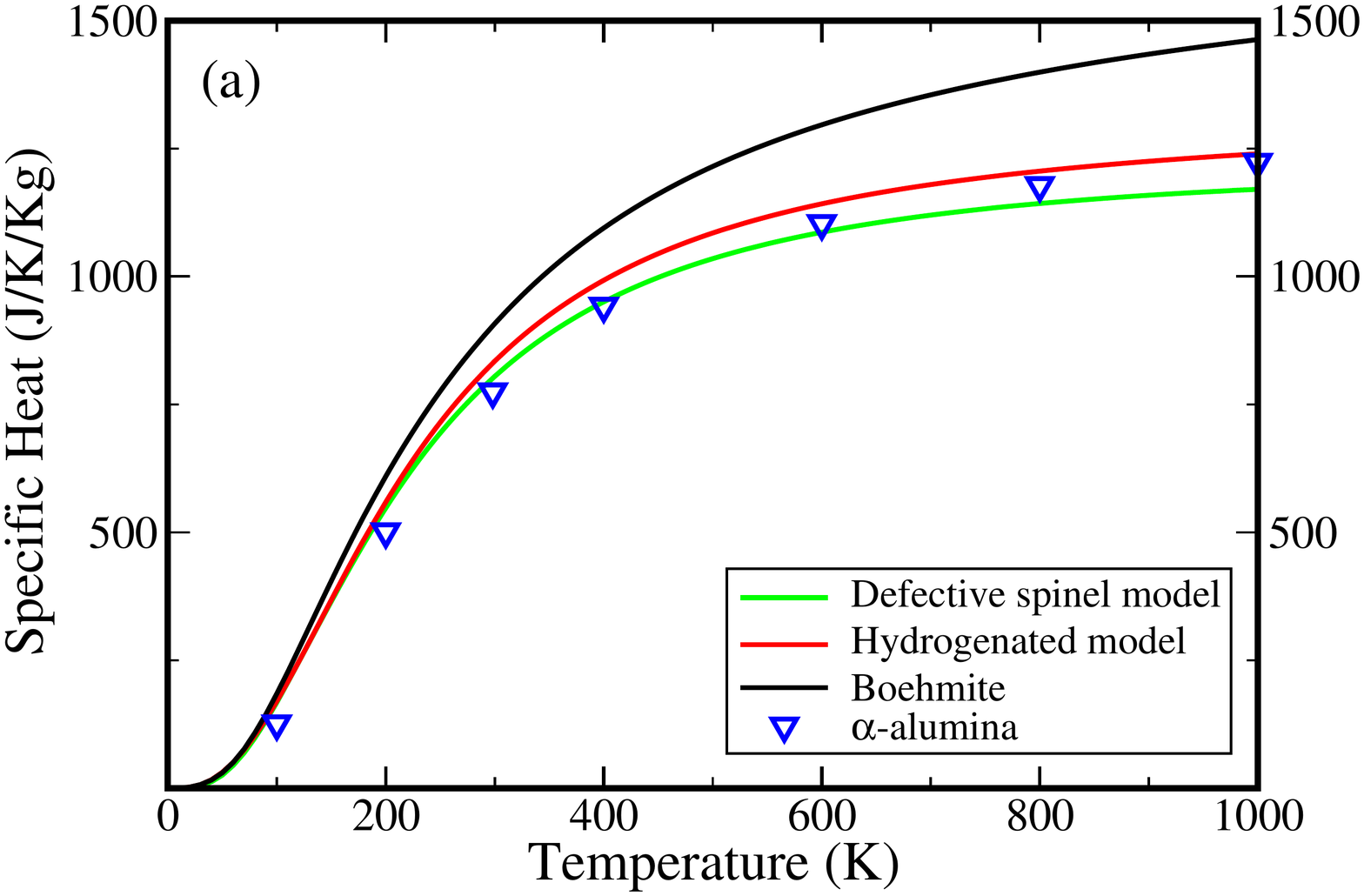}
\includegraphics[width=8.7 cm]{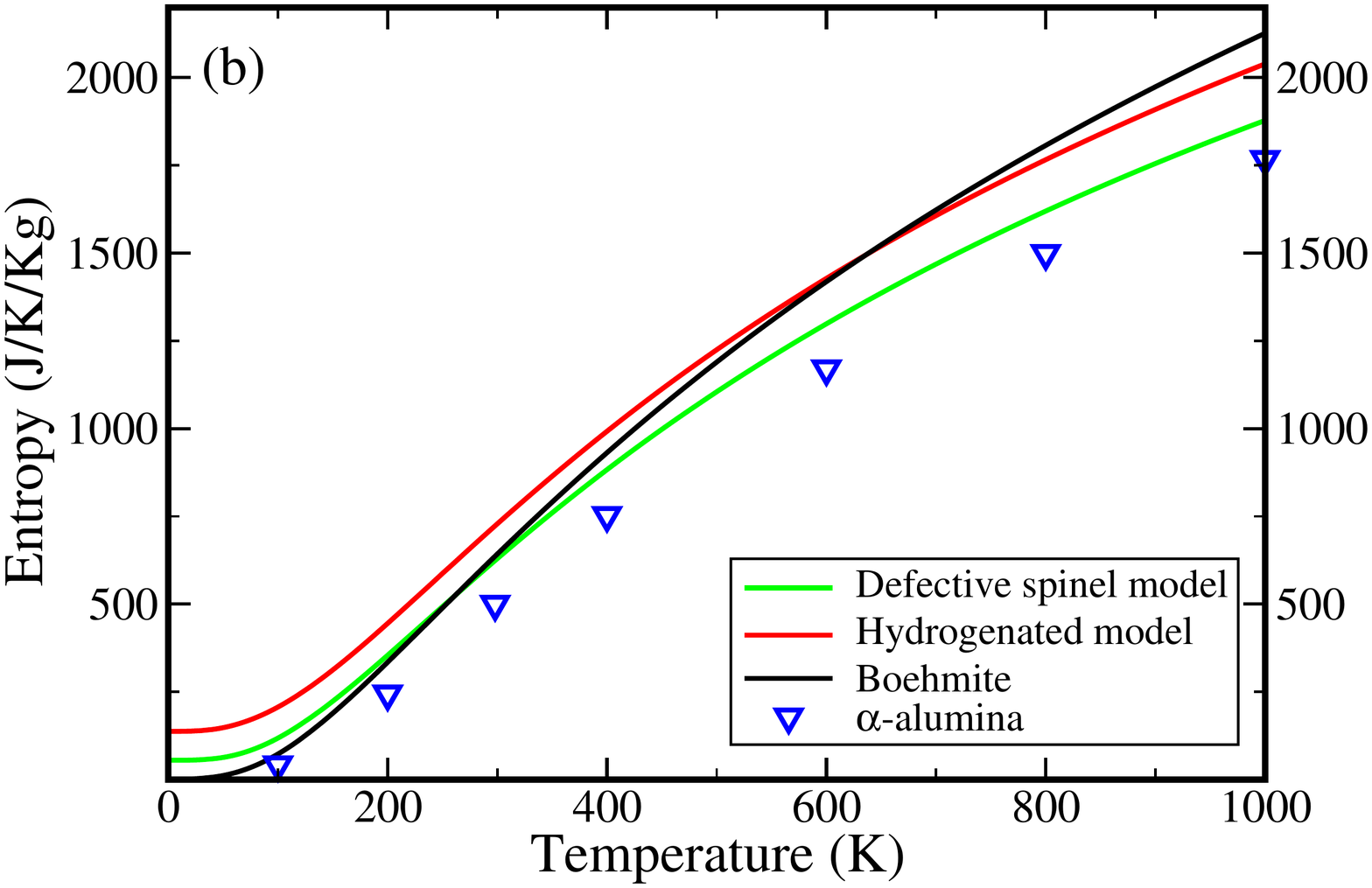}
\end{center}
\caption{\label{cv}Specific heat and entropy of $\gamma$-alumina and Boehmite in comparison with experimental data (symbols) for $\alpha$-alumina. Residual entropy of the spinel phases is included in panel (b).}
\end{figure}

Figure \ref{cv} shows the calculated temperature dependency of the specific heat and entropy for $\gamma$-alumina and Boehmite from 0 K to 1000 K, and the comparison with experimental data for $\alpha$-alumina.\cite{klug1987alumina} It is expected that the specific heat of $\gamma$-alumina lies in between those of Boehmite and $\alpha$-alumina, since $\gamma$-alumina is a transient phase during the decomposition of Boehmite into $\alpha$-alumina. As can be seen from Fig. \ref{cv}, Boehmite has a higher specific heat than any alumina. This is due to the hydrogenation, as hydrogenation can increase the specific heat and entropy of materials.\cite{kemp1937entropy} The hydrogenated spinel $\gamma$-alumina has a specific heat which lies in between that of the Boehmite and alumina. This can also be attributed to the hydrogenation. Generally, the $\alpha$-alumina has the lowest specific heat and entropy. Worth of notice is the fact that the defective spinel phase of $\gamma$-alumina has higher entropy than $\alpha$-alumina, but lower specific heat at temperatures above 500 K. This is because of the intrinsic vacancies in $\gamma$-alumina. These vacancies increase the disorder but also decrease the density of the material, resulting in higher entropy and lower specific heat values for $\gamma$-alumina in comparison to $\alpha$-alumina. To our knowledge, this is the first time that these differences in specific heat and entropy for $\gamma$-alumina and $\alpha$-alumina are reported. Such differences also distinguish the defective spinel phase from the hydrogenated spinel phase of $\gamma$-alumina.

\section{Summary}
We have investigated the electronic structure, dynamical stabilities, relative thermal stabilities and thermal properties of two competing phases of $\gamma$-alumina--the defectice spinel phase and the hydrogenated spinel phase--using periodic DFT calculations. The two $\gamma$-alumina phases exhibit similarities in electronic structure and the obtained crystalline parameters show good agreement with experimental data. Hybrid functional PBE0 was also used in calculations. Overall, the calculated equilibrium volume and bandgap of the hydrogenated spinel phase are close to the experimental data. The defective spinel phase is dynamically and thermally stable, and the obtained temperature at which its spontaneous formation from Boehmite occurs is in excellent agreement with experimental data. It is then shown that the methodologies followed in this study allow for accurate calculations of the transformation temperature. 

The defective spinel phase is found to be the ground state of $\gamma$-alumina. The hydrogenated spinel phase is also dynamically stable, but thermodynamically unstable with respect to the defective spinel phase and H$_2$O, as well as relative to defective spinel phase and Boehmite. This is in spite of the high entropy content of hydrogenated $\gamma$-alumina. The hydrogenated spinel structure is only a metastable phase during the decomposition of Boehmite above 753 K. However, dehydration of the metastable phase into the ground state is expected to be slow due to the low diffusion rate of H, which leaves hydrogen as a locked-in impurity in $\gamma$-alumina under conditions of normal temperature and pressure.

\begin{acknowledgments}
Financial support from the Swedish Foundation for Strategic Research (SSF, project ALUX) is gratefully acknowledged. Yunguo Li also thanks the Chinese Scholarship Council (CSC) for support. The computations were performed on resources
provided by the Swedish National Infrastructure for Computing (SNIC) at the National Supercomputer Center (NSC), Link\"oping, and at the PDC Center for High-performance Computing, Stockholm. 
\end{acknowledgments}

\appendix*
\section{}
\setcounter{table}{0}
\renewcommand\thetable{\Alph{section}.\arabic{table}}
 \begin{table}
 \caption{Crystallographic data (unit cell vectors and atomic coordinates) for computed defective spinel model of $\gamma$-alumina after
finding the symmetry with a tolerance of 0.001 \AA . Al(Td) and Al(Oh) indicate the TETRA Coord. and
OCT Coord. Al atoms, respectively.\label{table1}}
 \begin{ruledtabular}
 
\begin{tabular} {c c c c}
Space group           &                   & $\mathit{C2/m}$         &                       \\
$\mathit{a}$(\AA)    &16.71700                   & 0.03249                    & 0.02297                       \\
$\mathit{b}$(\AA)    &  2.79551                  & 4.90401                     &-0.01061                       \\
$\mathit{c}$(\AA)     &2.79551                   & 1.62467                      & 4.62701                      \\
\hline
 Sites  & $\mathit{x/a}$  &$\mathit{y/b}$   &$\mathit{z/c}$                      \\
Al1(Td)   &	0.00111  &	0.00835 &	0.00835		\\
Al2(Td)   &	0.08586 &	0.24425 &	0.24425		\\
Al3(Oh)  &	0.20833 &	0.12500 &	0.62500              		\\
Al4(Oh)   &0.20833 &	0.62500 & 0.12500	            	\\
Al5(Oh)  &	0.04797 &	0.63380  &	0.63380	            	\\
Al6(Td)   &	0.33081&	0.00575 &	0.00575		\\
Al7(Td)   &	0.41555&	0.24165 &	0.24165		\\
Al8(Oh)  &	0.54743&	0.61937 &	0.61937		\\
Al9(Oh)  &	0.54835&	0.10843 &	0.61945		\\
Al10(Oh)&	0.54835&	0.61945 &	0.10843		\\
Al11(Oh) &	0.36869&	0.6162   &	0.61620	           	\\
Al12(Td) &	0.66055&	0.00577&	0.00577		\\
Al13(Td) &	0.75611&	0.24423&	0.24423		\\
Al14(Oh)&	0.86923&	0.63063&	0.63063		\\
Al15(Oh)&	0.86832&	0.14157&	0.63055		\\
Al16(Oh)&	0.86832&	0.63055&	0.14157		\\
O1	    &	0.12494 &	0.38504&	0.38504		\\
O2	    &	0.12210 &	0.86484&	0.36667		\\
O3	    &	0.12210 &	0.36667&	0.86484	      \\
O4	    &	0.28748&	0.36683&	0.36683		\\
O5	    &	0.29457&	0.38516&	0.88333		\\
O6	    &	0.29172&	0.86496&	0.86496		\\
O7	    &	0.29457&	0.88333&	0.38516		\\
O8	    &	0.12919&	0.88317&	0.88317		\\
O9	    &	0.46068&	0.37775&	0.37775		\\
O10	    &	0.46282&	0.83823&	0.38111		\\
O11	    &	0.46282&	0.38111&	0.83823		\\
O12	    &	0.62131&	0.37578&	0.37578		\\
O13	    &	0.61436&	0.38672&	0.88202		\\
O14	    &	0.61440	      &	0.88271&	0.88271	      \\
O15	    &	0.61436&	0.88202&	0.38672		\\
O16	    &	0.46653&	0.86300	  &	0.86300	            	\\
O17	    &	0.80226&	0.36729&	0.36729		\\
O18	    &	0.80231&	0.86328&	0.36798		\\
O19	    &	0.80231&	0.36798&	0.86328		\\
O20	    &	0.95014&	0.38701&	0.38701		\\
O21	    &	0.95385&	0.41177&	0.86889		\\
O22	    &	0.95599&	0.87225&	0.87225		\\
O23	    &	0.95385&	0.86889&	0.41177		\\
O24	    &	0.79536&	0.87422&	0.87422		\\

 \end{tabular}
 \end{ruledtabular}
 \end{table}

\begin{table}
 \caption{Crystallographic data (unit cell vectors and atomic coordinates) for computed hydrogenated spinel model of $\gamma$-alumina after
finding the symmetry with a tolerance of 0.001 \AA. Al(Td) and Al(Oh) indicate the TETRA Coord. and
OCT Coord. Al atoms, respectively.\label{table2}}
 \begin{ruledtabular}
\begin{tabular} {c c c c}
Space group           &                   & $\mathit{P1}$         &                       \\
$\mathit{a}$(\AA)    & 16.66232       & -0.00004                    & 0.00014      \\
$\mathit{b}$(\AA)    &  2.74065      & 4.93336                      &-0.00002                       \\
$\mathit{c}$(\AA)     & 2.60751  &1.63688    & 4.62266                 \\
\hline
 Sites  & $\mathit{x/a}$  &$\mathit{y/b}$   &$\mathit{z/c}$                      \\
Al1(Td)	&	0.00421    &	0.00373	&	0.01219	\\
Al2(Td)	&	0.33755	&	0.00373	&	0.01219	\\
Al3(Td)	&	0.67088	&	0.00373	&	0.01219	\\
Al4(Td)	&	0.07992	&	0.26237	&	0.23932	\\
Al5(Td)	&	0.41325	&	0.26237	&	0.23932	\\
Al6(Td)	&	0.74659	&	0.26237	&	0.23932	\\
Al7(Oh)	&	0.20472	&	0.65726	&	0.13384	\\
Al8(Oh)	&	0.53806	&	0.65726	&	0.13384	\\
Al9(Oh)	&	0.87139	&	0.65726	&	0.13384	\\
Al10(Oh)	&	0.20560	&	0.13818    &	0.61635	\\
Al11(Oh)	&	0.53893	&	0.13818	&	0.61635	\\
Al12(Oh)	&	0.87226	&	0.13818	&	0.61635	\\
Al13(Oh)	&	0.04476	&	0.65727	&	0.61372	\\
Al14(Oh)	&	0.37809	&	0.65727	&	0.61372	\\
Al15(Oh)	&	0.71143	&	0.65727	&	0.61372	\\
O1	&	0.12403	&	0.40531	&	0.37165	\\
O2	&	0.45737	&	0.40531	&	0.37165	\\
O3	&	0.79070	&	0.40531	&	0.37165	\\
O4	&	0.12160	&	0.38401	&	0.85264	\\
O5	&	0.45494	&	0.38401	&	0.85264	\\
O6	&	0.78827	&	0.38401	&	0.85264	\\
O7	&	0.12342	&	0.88085	&	0.36983	\\
O8	&	0.45676	&	0.88085	&	0.36983	\\
O9	&	0.79009	&	0.88085	&	0.36983	\\
O10	&	0.28436	&	0.38401	&	0.36437	\\
O11	&	0.61769	&	0.38401	&	0.36437	\\
O12	&	0.95103	&	0.38401	&	0.36437	\\
O13	&	0.29234	&	0.89439    &	0.39288	\\
O14	&	0.62567     &	0.89439	&	0.39288	\\
O15	&	0.95900	&	0.89439	&	0.39288	\\
O16	&	0.29050	&	0.88034	&	0.87107	\\
O17	&	0.62384	&	0.88034	&	0.87107	\\
O18	&	0.95717	&	0.88034	&	0.87107	\\
O19	&	0.28759	&	0.40755	&	0.86231	\\
O20	&	0.62092	&	0.40755    &	0.86231	\\
O21	&	0.95425	&	0.40755	&	0.86231	\\
O22	&	0.13111	&	0.89439	&	0.87658	\\
O23	&	0.46444	&	0.89439	&	0.87658	\\
O24	&	0.79777	&	0.89439	&	0.87658	\\
H1	&	0.24718	&	0.47105	&	0.74109	\\
H2	&	0.58052	&	0.47105	&	0.74109	\\
H3	&	0.91385&	0.47105	&	0.74109\\

 \end{tabular}
 \end{ruledtabular}
 \end{table}

%

%\bibliography{aip}

\begin{thebibliography}{75}%
\makeatletter
\providecommand \@ifxundefined [1]{%
 \@ifx{#1\undefined}
}%
\providecommand \@ifnum [1]{%
 \ifnum #1\expandafter \@firstoftwo
 \else \expandafter \@secondoftwo
 \fi
}%
\providecommand \@ifx [1]{%
 \ifx #1\expandafter \@firstoftwo
 \else \expandafter \@secondoftwo
 \fi
}%
\providecommand \natexlab [1]{#1}%
\providecommand \enquote  [1]{``#1''}%
\providecommand \bibnamefont  [1]{#1}%
\providecommand \bibfnamefont [1]{#1}%
\providecommand \citenamefont [1]{#1}%
\providecommand \href@noop [0]{\@secondoftwo}%
\providecommand \href [0]{\begingroup \@sanitize@url \@href}%
\providecommand \@href[1]{\@@startlink{#1}\@@href}%
\providecommand \@@href[1]{\endgroup#1\@@endlink}%
\providecommand \@sanitize@url [0]{\catcode `\\12\catcode `\$12\catcode
  `\&12\catcode `\#12\catcode `\^12\catcode `\_12\catcode `\%12\relax}%
\providecommand \@@startlink[1]{}%
\providecommand \@@endlink[0]{}%
\providecommand \url  [0]{\begingroup\@sanitize@url \@url }%
\providecommand \@url [1]{\endgroup\@href {#1}{\urlprefix }}%
\providecommand \urlprefix  [0]{URL }%
\providecommand \Eprint [0]{\href }%
\providecommand \doibase [0]{http://dx.doi.org/}%
\providecommand \selectlanguage [0]{\@gobble}%
\providecommand \bibinfo  [0]{\@secondoftwo}%
\providecommand \bibfield  [0]{\@secondoftwo}%
\providecommand \translation [1]{[#1]}%
\providecommand \BibitemOpen [0]{}%
\providecommand \bibitemStop [0]{}%
\providecommand \bibitemNoStop [0]{.\EOS\space}%
\providecommand \EOS [0]{\spacefactor3000\relax}%
\providecommand \BibitemShut  [1]{\csname bibitem#1\endcsname}%
\let\auto@bib@innerbib\@empty
%</preamble>
\bibitem [{\citenamefont {D{\"o}rre}\ and\ \citenamefont
  {H{\"u}bner}(1984)}]{dorre1984alumina}%
  \BibitemOpen
  \bibfield  {author} {\bibinfo {author} {\bibfnamefont {E.}~\bibnamefont
  {D{\"o}rre}}\ and\ \bibinfo {author} {\bibfnamefont {H.}~\bibnamefont
  {H{\"u}bner}},\ }\href@noop {} {\emph {\bibinfo {title} {Alumina: processing,
  properties, and applications}}}\ (\bibinfo  {publisher} {Springer-Verlag
  Berlin},\ \bibinfo {year} {1984})\BibitemShut {NoStop}%
\bibitem [{\citenamefont {Hart}(1990)}]{hart1990alumina}%
  \BibitemOpen
  \bibfield  {author} {\bibinfo {author} {\bibfnamefont {L.~D.}\ \bibnamefont
  {Hart}},\ }\href@noop {} {\emph {\bibinfo {title} {Alumina chemicals}}}\
  (\bibinfo  {publisher} {Columbus, OH (USA); American Ceramic Society Inc.},\
  \bibinfo {year} {1990})\BibitemShut {NoStop}%
\bibitem [{\citenamefont {Zhou}\ and\ \citenamefont {Snyder}(1991)}]{zhou1991}%
  \BibitemOpen
  \bibfield  {author} {\bibinfo {author} {\bibfnamefont {R.-S.}\ \bibnamefont
  {Zhou}}\ and\ \bibinfo {author} {\bibfnamefont {R.~L.}\ \bibnamefont
  {Snyder}},\ }\href@noop {} {\bibfield  {journal} {\bibinfo  {journal} {Acta
  Crystallogr., Sect. B}\ }\textbf {\bibinfo {volume} {47}},\ \bibinfo {pages}
  {617} (\bibinfo {year} {1991})}\BibitemShut {NoStop}%
\bibitem [{\citenamefont {Levin}\ and\ \citenamefont
  {Brandon}(1998)}]{jace1995}%
  \BibitemOpen
  \bibfield  {author} {\bibinfo {author} {\bibfnamefont {I.}~\bibnamefont
  {Levin}}\ and\ \bibinfo {author} {\bibfnamefont {D.}~\bibnamefont
  {Brandon}},\ }\href@noop {} {\bibfield  {journal} {\bibinfo  {journal} {J.
  Am. Ceram. Soc.}\ }\textbf {\bibinfo {volume} {81}},\ \bibinfo {pages} {1995}
  (\bibinfo {year} {1998})}\BibitemShut {NoStop}%
\bibitem [{\citenamefont {Kn\"ozinger}\ and\ \citenamefont
  {Ratnasamy}(1978)}]{catalrev1978}%
  \BibitemOpen
  \bibfield  {author} {\bibinfo {author} {\bibfnamefont {H.}~\bibnamefont
  {Kn\"ozinger}}\ and\ \bibinfo {author} {\bibfnamefont {P.}~\bibnamefont
  {Ratnasamy}},\ }\href@noop {} {\bibfield  {journal} {\bibinfo  {journal}
  {Cat. Rev. - Sci. Eng.}\ }\textbf {\bibinfo {volume} {17}},\ \bibinfo {pages}
  {31} (\bibinfo {year} {1978})}\BibitemShut {NoStop}%
\bibitem [{\citenamefont {Hellman}\ and\ \citenamefont
  {Gr{\"o}nbeck}(2008)}]{hellman2008activation}%
  \BibitemOpen
  \bibfield  {author} {\bibinfo {author} {\bibfnamefont {A.}~\bibnamefont
  {Hellman}}\ and\ \bibinfo {author} {\bibfnamefont {H.}~\bibnamefont
  {Gr{\"o}nbeck}},\ }\href@noop {} {\bibfield  {journal} {\bibinfo  {journal}
  {Phys. Rev. Lett.}\ }\textbf {\bibinfo {volume} {100}},\ \bibinfo {pages}
  {116801} (\bibinfo {year} {2008})}\BibitemShut {NoStop}%
\bibitem [{\citenamefont {Scott}\ \emph {et~al.}(1987)\citenamefont {Scott},
  \citenamefont {Budge}, \citenamefont {Rheingold},\ and\ \citenamefont
  {Gates}}]{scott1987molecular}%
  \BibitemOpen
  \bibfield  {author} {\bibinfo {author} {\bibfnamefont {J.}~\bibnamefont
  {Scott}}, \bibinfo {author} {\bibfnamefont {J.}~\bibnamefont {Budge}},
  \bibinfo {author} {\bibfnamefont {A.}~\bibnamefont {Rheingold}}, \ and\
  \bibinfo {author} {\bibfnamefont {B.}~\bibnamefont {Gates}},\ }\href@noop {}
  {\bibfield  {journal} {\bibinfo  {journal} {J. Amer. Chem. Soc.}\ }\textbf
  {\bibinfo {volume} {109}},\ \bibinfo {pages} {7736} (\bibinfo {year}
  {1987})}\BibitemShut {NoStop}%
\bibitem [{\citenamefont {Wang}\ \emph {et~al.}(2004)\citenamefont {Wang},
  \citenamefont {Borisevich}, \citenamefont {Rashkeev}, \citenamefont
  {Glazoff}, \citenamefont {Sohlberg}, \citenamefont {Pennycook},\ and\
  \citenamefont {Pantelides}}]{wang2004dopants}%
  \BibitemOpen
  \bibfield  {author} {\bibinfo {author} {\bibfnamefont {S.}~\bibnamefont
  {Wang}}, \bibinfo {author} {\bibfnamefont {A.~Y.}\ \bibnamefont
  {Borisevich}}, \bibinfo {author} {\bibfnamefont {S.~N.}\ \bibnamefont
  {Rashkeev}}, \bibinfo {author} {\bibfnamefont {M.~V.}\ \bibnamefont
  {Glazoff}}, \bibinfo {author} {\bibfnamefont {K.}~\bibnamefont {Sohlberg}},
  \bibinfo {author} {\bibfnamefont {S.~J.}\ \bibnamefont {Pennycook}}, \ and\
  \bibinfo {author} {\bibfnamefont {S.~T.}\ \bibnamefont {Pantelides}},\
  }\href@noop {} {\bibfield  {journal} {\bibinfo  {journal} {Nat. Mater.}\
  }\textbf {\bibinfo {volume} {3}},\ \bibinfo {pages} {143} (\bibinfo {year}
  {2004})}\BibitemShut {NoStop}%
\bibitem [{\citenamefont {Stumpf}\ \emph {et~al.}(1950)\citenamefont {Stumpf},
  \citenamefont {Russell}, \citenamefont {Newsome},\ and\ \citenamefont
  {Tucker}}]{stumpf1950thermal}%
  \BibitemOpen
  \bibfield  {author} {\bibinfo {author} {\bibfnamefont {H.}~\bibnamefont
  {Stumpf}}, \bibinfo {author} {\bibfnamefont {A.~S.}\ \bibnamefont {Russell}},
  \bibinfo {author} {\bibfnamefont {J.}~\bibnamefont {Newsome}}, \ and\
  \bibinfo {author} {\bibfnamefont {C.}~\bibnamefont {Tucker}},\ }\href@noop {}
  {\bibfield  {journal} {\bibinfo  {journal} {Ind. Eng. Chem.}\ }\textbf
  {\bibinfo {volume} {42}},\ \bibinfo {pages} {1398} (\bibinfo {year}
  {1950})}\BibitemShut {NoStop}%
\bibitem [{\citenamefont {Ushakov}\ and\ \citenamefont
  {Moroz}(1984)}]{ushakov1984structure}%
  \BibitemOpen
  \bibfield  {author} {\bibinfo {author} {\bibfnamefont {V.}~\bibnamefont
  {Ushakov}}\ and\ \bibinfo {author} {\bibfnamefont {E.}~\bibnamefont
  {Moroz}},\ }\href@noop {} {\bibfield  {journal} {\bibinfo  {journal} {React.
  Kinet. Catal. Lett.}\ }\textbf {\bibinfo {volume} {24}},\ \bibinfo {pages}
  {113} (\bibinfo {year} {1984})}\BibitemShut {NoStop}%
\bibitem [{\citenamefont {Tsyganenko}\ \emph {et~al.}(1990)\citenamefont
  {Tsyganenko}, \citenamefont {Smirnov}, \citenamefont {Rzhevskij},\ and\
  \citenamefont {Mardilovich}}]{tsyganenko1990infrared}%
  \BibitemOpen
  \bibfield  {author} {\bibinfo {author} {\bibfnamefont {A.}~\bibnamefont
  {Tsyganenko}}, \bibinfo {author} {\bibfnamefont {K.}~\bibnamefont {Smirnov}},
  \bibinfo {author} {\bibfnamefont {A.}~\bibnamefont {Rzhevskij}}, \ and\
  \bibinfo {author} {\bibfnamefont {P.}~\bibnamefont {Mardilovich}},\
  }\href@noop {} {\bibfield  {journal} {\bibinfo  {journal} {Mater. Chem.
  Phys.}\ }\textbf {\bibinfo {volume} {26}},\ \bibinfo {pages} {35} (\bibinfo
  {year} {1990})}\BibitemShut {NoStop}%
\bibitem [{\citenamefont {Sohlberg}\ \emph {et~al.}(1999)\citenamefont
  {Sohlberg}, \citenamefont {Pennycook},\ and\ \citenamefont
  {Pantelides}}]{sohlberg1999hydrogen}%
  \BibitemOpen
  \bibfield  {author} {\bibinfo {author} {\bibfnamefont {K.}~\bibnamefont
  {Sohlberg}}, \bibinfo {author} {\bibfnamefont {S.~J.}\ \bibnamefont
  {Pennycook}}, \ and\ \bibinfo {author} {\bibfnamefont {S.~T.}\ \bibnamefont
  {Pantelides}},\ }\href@noop {} {\bibfield  {journal} {\bibinfo  {journal} {J.
  Am. Chem. Soc.}\ }\textbf {\bibinfo {volume} {121}},\ \bibinfo {pages} {7493}
  (\bibinfo {year} {1999})}\BibitemShut {NoStop}%
\bibitem [{\citenamefont {Wolverton}\ and\ \citenamefont
  {Hass}(2000)}]{wolverton2000phase}%
  \BibitemOpen
  \bibfield  {author} {\bibinfo {author} {\bibfnamefont {C.}~\bibnamefont
  {Wolverton}}\ and\ \bibinfo {author} {\bibfnamefont {K.}~\bibnamefont
  {Hass}},\ }\href@noop {} {\bibfield  {journal} {\bibinfo  {journal} {Phys.
  Rev. B}\ }\textbf {\bibinfo {volume} {63}},\ \bibinfo {pages} {024102}
  (\bibinfo {year} {2000})}\BibitemShut {NoStop}%
\bibitem [{\citenamefont {Ching}\ \emph {et~al.}(2008)\citenamefont {Ching},
  \citenamefont {Ouyang}, \citenamefont {Rulis},\ and\ \citenamefont
  {Yao}}]{PhysRevB.78.014106}%
  \BibitemOpen
  \bibfield  {author} {\bibinfo {author} {\bibfnamefont {W.~Y.}\ \bibnamefont
  {Ching}}, \bibinfo {author} {\bibfnamefont {L.}~\bibnamefont {Ouyang}},
  \bibinfo {author} {\bibfnamefont {P.}~\bibnamefont {Rulis}}, \ and\ \bibinfo
  {author} {\bibfnamefont {H.}~\bibnamefont {Yao}},\ }\href {\doibase
  10.1103/PhysRevB.78.014106} {\bibfield  {journal} {\bibinfo  {journal} {Phys.
  Rev. B}\ }\textbf {\bibinfo {volume} {78}},\ \bibinfo {pages} {014106}
  (\bibinfo {year} {2008})}\BibitemShut {NoStop}%
\bibitem [{\citenamefont {Paglia}\ \emph {et~al.}(2005)\citenamefont {Paglia},
  \citenamefont {Rohl}, \citenamefont {Buckley},\ and\ \citenamefont
  {Gale}}]{paglia2005determination}%
  \BibitemOpen
  \bibfield  {author} {\bibinfo {author} {\bibfnamefont {G.}~\bibnamefont
  {Paglia}}, \bibinfo {author} {\bibfnamefont {A.~L.}\ \bibnamefont {Rohl}},
  \bibinfo {author} {\bibfnamefont {C.~E.}\ \bibnamefont {Buckley}}, \ and\
  \bibinfo {author} {\bibfnamefont {J.~D.}\ \bibnamefont {Gale}},\ }\href@noop
  {} {\bibfield  {journal} {\bibinfo  {journal} {Phys. Rev. B}\ }\textbf
  {\bibinfo {volume} {71}},\ \bibinfo {pages} {224115} (\bibinfo {year}
  {2005})}\BibitemShut {NoStop}%
\bibitem [{\citenamefont {Jennison}\ \emph {et~al.}(2004)\citenamefont
  {Jennison}, \citenamefont {Schultz},\ and\ \citenamefont
  {Sullivan}}]{PhysRevB.69.041405}%
  \BibitemOpen
  \bibfield  {author} {\bibinfo {author} {\bibfnamefont {D.~R.}\ \bibnamefont
  {Jennison}}, \bibinfo {author} {\bibfnamefont {P.~A.}\ \bibnamefont
  {Schultz}}, \ and\ \bibinfo {author} {\bibfnamefont {J.~P.}\ \bibnamefont
  {Sullivan}},\ }\href {\doibase 10.1103/PhysRevB.69.041405} {\bibfield
  {journal} {\bibinfo  {journal} {Phys. Rev. B}\ }\textbf {\bibinfo {volume}
  {69}},\ \bibinfo {pages} {041405} (\bibinfo {year} {2004})}\BibitemShut
  {NoStop}%
\bibitem [{\citenamefont {Ferreira}\ \emph {et~al.}(2011)\citenamefont
  {Ferreira}, \citenamefont {Martins}, \citenamefont {Konstantinova},
  \citenamefont {Capaz}, \citenamefont {Souza}, \citenamefont {Chiaro},\ and\
  \citenamefont {Leit{\~a}o}}]{ferreira2011direct}%
  \BibitemOpen
  \bibfield  {author} {\bibinfo {author} {\bibfnamefont {A.~R.}\ \bibnamefont
  {Ferreira}}, \bibinfo {author} {\bibfnamefont {M.~J.}\ \bibnamefont
  {Martins}}, \bibinfo {author} {\bibfnamefont {E.}~\bibnamefont
  {Konstantinova}}, \bibinfo {author} {\bibfnamefont {R.~B.}\ \bibnamefont
  {Capaz}}, \bibinfo {author} {\bibfnamefont {W.~F.}\ \bibnamefont {Souza}},
  \bibinfo {author} {\bibfnamefont {S.~S.~X.}\ \bibnamefont {Chiaro}}, \ and\
  \bibinfo {author} {\bibfnamefont {A.~A.}\ \bibnamefont {Leit{\~a}o}},\
  }\href@noop {} {\bibfield  {journal} {\bibinfo  {journal} {J. Solid State
  Chem.}\ }\textbf {\bibinfo {volume} {184}},\ \bibinfo {pages} {1105}
  (\bibinfo {year} {2011})}\BibitemShut {NoStop}%
\bibitem [{\citenamefont {Trueba}\ and\ \citenamefont
  {Trasatti}(2005)}]{trueba2005gamma}%
  \BibitemOpen
  \bibfield  {author} {\bibinfo {author} {\bibfnamefont {M.}~\bibnamefont
  {Trueba}}\ and\ \bibinfo {author} {\bibfnamefont {S.~P.}\ \bibnamefont
  {Trasatti}},\ }\href@noop {} {\bibfield  {journal} {\bibinfo  {journal} {Eur.
  J. Inorg. Chem.}\ }\textbf {\bibinfo {volume} {2005}},\ \bibinfo {pages}
  {3393} (\bibinfo {year} {2005})}\BibitemShut {NoStop}%
\bibitem [{\citenamefont {Digne}\ \emph {et~al.}(2004)\citenamefont {Digne},
  \citenamefont {Sautet}, \citenamefont {Raybaud}, \citenamefont {Euzen},\ and\
  \citenamefont {Toulhoat}}]{digne2004use}%
  \BibitemOpen
  \bibfield  {author} {\bibinfo {author} {\bibfnamefont {M.}~\bibnamefont
  {Digne}}, \bibinfo {author} {\bibfnamefont {P.}~\bibnamefont {Sautet}},
  \bibinfo {author} {\bibfnamefont {P.}~\bibnamefont {Raybaud}}, \bibinfo
  {author} {\bibfnamefont {P.}~\bibnamefont {Euzen}}, \ and\ \bibinfo {author}
  {\bibfnamefont {H.}~\bibnamefont {Toulhoat}},\ }\href@noop {} {\bibfield
  {journal} {\bibinfo  {journal} {J. Catal.}\ }\textbf {\bibinfo {volume}
  {226}},\ \bibinfo {pages} {54} (\bibinfo {year} {2004})}\BibitemShut
  {NoStop}%
\bibitem [{\citenamefont {Hass}\ \emph {et~al.}(1998)\citenamefont {Hass},
  \citenamefont {Schneider}, \citenamefont {Curioni},\ and\ \citenamefont
  {Andreoni}}]{hass1998chemistry}%
  \BibitemOpen
  \bibfield  {author} {\bibinfo {author} {\bibfnamefont {K.~C.}\ \bibnamefont
  {Hass}}, \bibinfo {author} {\bibfnamefont {W.~F.}\ \bibnamefont {Schneider}},
  \bibinfo {author} {\bibfnamefont {A.}~\bibnamefont {Curioni}}, \ and\
  \bibinfo {author} {\bibfnamefont {W.}~\bibnamefont {Andreoni}},\ }\href@noop
  {} {\bibfield  {journal} {\bibinfo  {journal} {Science}\ }\textbf {\bibinfo
  {volume} {282}},\ \bibinfo {pages} {265} (\bibinfo {year}
  {1998})}\BibitemShut {NoStop}%
\bibitem [{\citenamefont {Dowden}(1950)}]{dowden195056}%
  \BibitemOpen
  \bibfield  {author} {\bibinfo {author} {\bibfnamefont {D.}~\bibnamefont
  {Dowden}},\ }\href@noop {} {\bibfield  {journal} {\bibinfo  {journal} {J.
  Chem. Soc. (Resumed)}\ ,\ \bibinfo {pages} {242}} (\bibinfo {year}
  {1950})}\BibitemShut {NoStop}%
\bibitem [{\citenamefont {Rashkeev}\ \emph {et~al.}(2007)\citenamefont
  {Rashkeev}, \citenamefont {Sohlberg}, \citenamefont {Zhuo},\ and\
  \citenamefont {Pantelides}}]{rashkeev2007hydrogen}%
  \BibitemOpen
  \bibfield  {author} {\bibinfo {author} {\bibfnamefont {S.~N.}\ \bibnamefont
  {Rashkeev}}, \bibinfo {author} {\bibfnamefont {K.~W.}\ \bibnamefont
  {Sohlberg}}, \bibinfo {author} {\bibfnamefont {S.}~\bibnamefont {Zhuo}}, \
  and\ \bibinfo {author} {\bibfnamefont {S.~T.}\ \bibnamefont {Pantelides}},\
  }\href@noop {} {\bibfield  {journal} {\bibinfo  {journal} {J. Phys. Chem. C}\
  }\textbf {\bibinfo {volume} {111}},\ \bibinfo {pages} {7175} (\bibinfo {year}
  {2007})}\BibitemShut {NoStop}%
\bibitem [{\citenamefont {Lippens}\ and\ \citenamefont
  {Steggerda}(1970)}]{lippens1970}%
  \BibitemOpen
  \bibfield  {author} {\bibinfo {author} {\bibfnamefont {B.~C.}\ \bibnamefont
  {Lippens}}\ and\ \bibinfo {author} {\bibfnamefont {J.~J.}\ \bibnamefont
  {Steggerda}},\ }\href@noop {} {\emph {\bibinfo {title} {Physical and Chemical
  Aspects of Adsorbents and Catalysts}}},\ edited by\ \bibinfo {editor}
  {\bibfnamefont {B.~G.}\ \bibnamefont {Linsen}}\ (\bibinfo  {publisher}
  {Academic Press},\ \bibinfo {address} {London},\ \bibinfo {year}
  {1970})\BibitemShut {NoStop}%
\bibitem [{\citenamefont {Sohlberg}\ \emph {et~al.}(2000)\citenamefont
  {Sohlberg}, \citenamefont {Pennycook},\ and\ \citenamefont
  {Pantelides}}]{sohlberg2000bulk}%
  \BibitemOpen
  \bibfield  {author} {\bibinfo {author} {\bibfnamefont {K.}~\bibnamefont
  {Sohlberg}}, \bibinfo {author} {\bibfnamefont {S.~J.}\ \bibnamefont
  {Pennycook}}, \ and\ \bibinfo {author} {\bibfnamefont {S.~T.}\ \bibnamefont
  {Pantelides}},\ }\href@noop {} {\bibfield  {journal} {\bibinfo  {journal}
  {Chem. Eng. Commun.}\ }\textbf {\bibinfo {volume} {181}},\ \bibinfo {pages}
  {107} (\bibinfo {year} {2000})}\BibitemShut {NoStop}%
\bibitem [{\citenamefont {Paglia}\ \emph {et~al.}(2004)\citenamefont {Paglia},
  \citenamefont {Buckley}, \citenamefont {Rohl}, \citenamefont {Hart},
  \citenamefont {Winter}, \citenamefont {Studer}, \citenamefont {Hunter},\ and\
  \citenamefont {Hanna}}]{paglia2004boehmite}%
  \BibitemOpen
  \bibfield  {author} {\bibinfo {author} {\bibfnamefont {G.}~\bibnamefont
  {Paglia}}, \bibinfo {author} {\bibfnamefont {C.~E.}\ \bibnamefont {Buckley}},
  \bibinfo {author} {\bibfnamefont {A.~L.}\ \bibnamefont {Rohl}}, \bibinfo
  {author} {\bibfnamefont {R.~D.}\ \bibnamefont {Hart}}, \bibinfo {author}
  {\bibfnamefont {K.}~\bibnamefont {Winter}}, \bibinfo {author} {\bibfnamefont
  {A.~J.}\ \bibnamefont {Studer}}, \bibinfo {author} {\bibfnamefont {B.~A.}\
  \bibnamefont {Hunter}}, \ and\ \bibinfo {author} {\bibfnamefont {J.~V.}\
  \bibnamefont {Hanna}},\ }\href@noop {} {\bibfield  {journal} {\bibinfo
  {journal} {Chem. Mater.}\ }\textbf {\bibinfo {volume} {16}},\ \bibinfo
  {pages} {220} (\bibinfo {year} {2004})}\BibitemShut {NoStop}%
\bibitem [{\citenamefont {Verwey}(1935)}]{verwey1935electrolytic}%
  \BibitemOpen
  \bibfield  {author} {\bibinfo {author} {\bibfnamefont {E.}~\bibnamefont
  {Verwey}},\ }\href@noop {} {\bibfield  {journal} {\bibinfo  {journal}
  {Physica}\ }\textbf {\bibinfo {volume} {2}},\ \bibinfo {pages} {1059}
  (\bibinfo {year} {1935})}\BibitemShut {NoStop}%
\bibitem [{\citenamefont {Jagodzinski}\ and\ \citenamefont
  {Saalfeld}(1958)}]{jagodzinski1958}%
  \BibitemOpen
  \bibfield  {author} {\bibinfo {author} {\bibfnamefont {H.~T.}\ \bibnamefont
  {Jagodzinski}}\ and\ \bibinfo {author} {\bibfnamefont {H.}~\bibnamefont
  {Saalfeld}},\ }\href@noop {} {\bibfield  {journal} {\bibinfo  {journal}
  {Zeitschrift f{\"u}r Kristallographie}\ }\textbf {\bibinfo {volume} {110}},\
  \bibinfo {pages} {197} (\bibinfo {year} {1958})}\BibitemShut {NoStop}%
\bibitem [{\citenamefont {Saalfeld}(1960)}]{saalfeld1960strukturen}%
  \BibitemOpen
  \bibfield  {author} {\bibinfo {author} {\bibfnamefont {H.}~\bibnamefont
  {Saalfeld}},\ }\href@noop {} {\bibfield  {journal} {\bibinfo  {journal}
  {Neues Jahrb Mineral Abh}\ }\textbf {\bibinfo {volume} {95}},\ \bibinfo
  {pages} {1} (\bibinfo {year} {1960})}\BibitemShut {NoStop}%
\bibitem [{\citenamefont {Lee}\ \emph {et~al.}(1997)\citenamefont {Lee},
  \citenamefont {Cheng}, \citenamefont {Heine},\ and\ \citenamefont
  {Klinowski}}]{lee1997distribution}%
  \BibitemOpen
  \bibfield  {author} {\bibinfo {author} {\bibfnamefont {M.-H.}\ \bibnamefont
  {Lee}}, \bibinfo {author} {\bibfnamefont {C.-F.}\ \bibnamefont {Cheng}},
  \bibinfo {author} {\bibfnamefont {V.}~\bibnamefont {Heine}}, \ and\ \bibinfo
  {author} {\bibfnamefont {J.}~\bibnamefont {Klinowski}},\ }\href@noop {}
  {\bibfield  {journal} {\bibinfo  {journal} {Chem. Phys. Lett.}\ }\textbf
  {\bibinfo {volume} {265}},\ \bibinfo {pages} {673} (\bibinfo {year}
  {1997})}\BibitemShut {NoStop}%
\bibitem [{\citenamefont {Pecharrom\~an}\ \emph {et~al.}(1999)\citenamefont
  {Pecharrom\~an}, \citenamefont {Sobrados}, \citenamefont {Iglesias},
  \citenamefont {Gonz\~alez Carre\~no},\ and\ \citenamefont
  {Sanz}}]{pecharroman1999thermal}%
  \BibitemOpen
  \bibfield  {author} {\bibinfo {author} {\bibfnamefont {C.}~\bibnamefont
  {Pecharrom\~an}}, \bibinfo {author} {\bibfnamefont {I.}~\bibnamefont
  {Sobrados}}, \bibinfo {author} {\bibfnamefont {J.}~\bibnamefont {Iglesias}},
  \bibinfo {author} {\bibfnamefont {T.}~\bibnamefont {Gonz\~alez Carre\~no}}, \
  and\ \bibinfo {author} {\bibfnamefont {J.}~\bibnamefont {Sanz}},\ }\href@noop
  {} {\bibfield  {journal} {\bibinfo  {journal} {J. Phys. Chem. B}\ }\textbf
  {\bibinfo {volume} {103}},\ \bibinfo {pages} {6160} (\bibinfo {year}
  {1999})}\BibitemShut {NoStop}%
\bibitem [{\citenamefont {John}\ \emph {et~al.}(1983)\citenamefont {John},
  \citenamefont {Alma},\ and\ \citenamefont {Hays}}]{john1983characterization}%
  \BibitemOpen
  \bibfield  {author} {\bibinfo {author} {\bibfnamefont {C.}~\bibnamefont
  {John}}, \bibinfo {author} {\bibfnamefont {N.}~\bibnamefont {Alma}}, \ and\
  \bibinfo {author} {\bibfnamefont {G.}~\bibnamefont {Hays}},\ }\href@noop {}
  {\bibfield  {journal} {\bibinfo  {journal} {Appl. Catal.}\ }\textbf {\bibinfo
  {volume} {6}},\ \bibinfo {pages} {341} (\bibinfo {year} {1983})}\BibitemShut
  {NoStop}%
\bibitem [{\citenamefont {Mo}\ \emph {et~al.}(1997)\citenamefont {Mo},
  \citenamefont {Xu},\ and\ \citenamefont {Ching}}]{mo1997electronic}%
  \BibitemOpen
  \bibfield  {author} {\bibinfo {author} {\bibfnamefont {S.-D.}\ \bibnamefont
  {Mo}}, \bibinfo {author} {\bibfnamefont {Y.-N.}\ \bibnamefont {Xu}}, \ and\
  \bibinfo {author} {\bibfnamefont {W.-Y.}\ \bibnamefont {Ching}},\ }\href@noop
  {} {\bibfield  {journal} {\bibinfo  {journal} {J. Am. Ceram. Soc.}\ }\textbf
  {\bibinfo {volume} {80}},\ \bibinfo {pages} {1193} (\bibinfo {year}
  {1997})}\BibitemShut {NoStop}%
\bibitem [{\citenamefont {Sun}\ \emph {et~al.}(2006)\citenamefont {Sun},
  \citenamefont {Nelson},\ and\ \citenamefont {Adjaye}}]{sun2006examination}%
  \BibitemOpen
  \bibfield  {author} {\bibinfo {author} {\bibfnamefont {M.}~\bibnamefont
  {Sun}}, \bibinfo {author} {\bibfnamefont {A.~E.}\ \bibnamefont {Nelson}}, \
  and\ \bibinfo {author} {\bibfnamefont {J.}~\bibnamefont {Adjaye}},\
  }\href@noop {} {\bibfield  {journal} {\bibinfo  {journal} {J. Phys. Chem. B}\
  }\textbf {\bibinfo {volume} {110}},\ \bibinfo {pages} {2310} (\bibinfo {year}
  {2006})}\BibitemShut {NoStop}%
\bibitem [{\citenamefont {Paglia}\ \emph {et~al.}(2006)\citenamefont {Paglia},
  \citenamefont {Buckley},\ and\ \citenamefont {Rohl}}]{paglia2006comment}%
  \BibitemOpen
  \bibfield  {author} {\bibinfo {author} {\bibfnamefont {G.}~\bibnamefont
  {Paglia}}, \bibinfo {author} {\bibfnamefont {C.}~\bibnamefont {Buckley}}, \
  and\ \bibinfo {author} {\bibfnamefont {A.~L.}\ \bibnamefont {Rohl}},\
  }\href@noop {} {\bibfield  {journal} {\bibinfo  {journal} {J. Phys. Chem. B}\
  }\textbf {\bibinfo {volume} {110}},\ \bibinfo {pages} {20721} (\bibinfo
  {year} {2006})}\BibitemShut {NoStop}%
\bibitem [{\citenamefont {Digne}\ \emph {et~al.}(2006)\citenamefont {Digne},
  \citenamefont {Raybaud}, \citenamefont {Sautet}, \citenamefont {Rebours},
  \citenamefont {Toulhoat}, \citenamefont {Paglia}, \citenamefont {Buckley},
  \citenamefont {Rohl}, \citenamefont {Nelson},\ and\ \citenamefont
  {M.}}]{digne2006comment}%
  \BibitemOpen
  \bibfield  {author} {\bibinfo {author} {\bibfnamefont {M.}~\bibnamefont
  {Digne}}, \bibinfo {author} {\bibfnamefont {P.}~\bibnamefont {Raybaud}},
  \bibinfo {author} {\bibfnamefont {P.}~\bibnamefont {Sautet}}, \bibinfo
  {author} {\bibfnamefont {B.}~\bibnamefont {Rebours}}, \bibinfo {author}
  {\bibfnamefont {H.}~\bibnamefont {Toulhoat}}, \bibinfo {author}
  {\bibfnamefont {G.}~\bibnamefont {Paglia}}, \bibinfo {author} {\bibfnamefont
  {C.}~\bibnamefont {Buckley}}, \bibinfo {author} {\bibfnamefont {A.~L.}\
  \bibnamefont {Rohl}}, \bibinfo {author} {\bibfnamefont {A.~E.}\ \bibnamefont
  {Nelson}}, \ and\ \bibinfo {author} {\bibfnamefont {S.}~\bibnamefont {M.}},\
  }\href@noop {} {\bibfield  {journal} {\bibinfo  {journal} {J. Phys. Chem. B}\
  }\textbf {\bibinfo {volume} {110}},\ \bibinfo {pages} {20719} (\bibinfo
  {year} {2006})}\BibitemShut {NoStop}%
\bibitem [{\citenamefont {Alvarez}\ \emph {et~al.}(1992)\citenamefont
  {Alvarez}, \citenamefont {Sanz}, \citenamefont {Capit{\'a}n},\ and\
  \citenamefont {Odriozola}}]{alvarez1992molecular}%
  \BibitemOpen
  \bibfield  {author} {\bibinfo {author} {\bibfnamefont {L.~J.}\ \bibnamefont
  {Alvarez}}, \bibinfo {author} {\bibfnamefont {J.~F.}\ \bibnamefont {Sanz}},
  \bibinfo {author} {\bibfnamefont {M.~J.}\ \bibnamefont {Capit{\'a}n}}, \ and\
  \bibinfo {author} {\bibfnamefont {J.~A.}\ \bibnamefont {Odriozola}},\
  }\href@noop {} {\bibfield  {journal} {\bibinfo  {journal} {Chem. Phys.
  Lett.}\ }\textbf {\bibinfo {volume} {192}},\ \bibinfo {pages} {463} (\bibinfo
  {year} {1992})}\BibitemShut {NoStop}%
\bibitem [{\citenamefont {de~Boer}\ \emph {et~al.}(1952)\citenamefont
  {de~Boer}, \citenamefont {Houben},\ and\ \citenamefont
  {Terpstra}}]{de1952proceedings}%
  \BibitemOpen
  \bibfield  {author} {\bibinfo {author} {\bibfnamefont {J.}~\bibnamefont
  {de~Boer}}, \bibinfo {author} {\bibfnamefont {G.}~\bibnamefont {Houben}}, \
  and\ \bibinfo {author} {\bibfnamefont {R.}~\bibnamefont {Terpstra}},\
  }\href@noop {} {\enquote {\bibinfo {title} {Proceedings of the international
  symposium on the reactivity of solids},}\ } (\bibinfo {year}
  {1952})\BibitemShut {NoStop}%
\bibitem [{\citenamefont {Maciver}\ \emph {et~al.}(1963)\citenamefont
  {Maciver}, \citenamefont {Tobin},\ and\ \citenamefont
  {Barth}}]{maciver1963catalytic}%
  \BibitemOpen
  \bibfield  {author} {\bibinfo {author} {\bibfnamefont {D.}~\bibnamefont
  {Maciver}}, \bibinfo {author} {\bibfnamefont {H.}~\bibnamefont {Tobin}}, \
  and\ \bibinfo {author} {\bibfnamefont {R.}~\bibnamefont {Barth}},\
  }\href@noop {} {\bibfield  {journal} {\bibinfo  {journal} {J. Catal.}\
  }\textbf {\bibinfo {volume} {2}},\ \bibinfo {pages} {485} (\bibinfo {year}
  {1963})}\BibitemShut {NoStop}%
\bibitem [{\citenamefont {L{\'e}onard}\ \emph {et~al.}(1969)\citenamefont
  {L{\'e}onard}, \citenamefont {Semaille},\ and\ \citenamefont
  {Fripiat}}]{leonard1969structure}%
  \BibitemOpen
  \bibfield  {author} {\bibinfo {author} {\bibfnamefont {A.}~\bibnamefont
  {L{\'e}onard}}, \bibinfo {author} {\bibfnamefont {P.}~\bibnamefont
  {Semaille}}, \ and\ \bibinfo {author} {\bibfnamefont {J.}~\bibnamefont
  {Fripiat}},\ }in\ \href@noop {} {\emph {\bibinfo {booktitle} {Proc. Br.
  Ceram. Soc}}},\ Vol.\ \bibinfo {volume} {103}\ (\bibinfo {year} {1969})\ p.\
  \bibinfo {pages} {103}\BibitemShut {NoStop}%
\bibitem [{\citenamefont {Pearson}(1971)}]{pearson1971wide}%
  \BibitemOpen
  \bibfield  {author} {\bibinfo {author} {\bibfnamefont {R.}~\bibnamefont
  {Pearson}},\ }\href@noop {} {\bibfield  {journal} {\bibinfo  {journal} {J.
  Catal.}\ }\textbf {\bibinfo {volume} {23}},\ \bibinfo {pages} {388} (\bibinfo
  {year} {1971})}\BibitemShut {NoStop}%
\bibitem [{\citenamefont {Soled}(1983)}]{Soled1983252}%
  \BibitemOpen
  \bibfield  {author} {\bibinfo {author} {\bibfnamefont {S.}~\bibnamefont
  {Soled}},\ }\href@noop {} {\bibfield  {journal} {\bibinfo  {journal} {J.
  Catal.}\ }\textbf {\bibinfo {volume} {81}},\ \bibinfo {pages} {252 }
  (\bibinfo {year} {1983})}\BibitemShut {NoStop}%
\bibitem [{\citenamefont {White}\ \emph {et~al.}(1989)\citenamefont {White},
  \citenamefont {McHargue}, \citenamefont {Sklad}, \citenamefont {Boatner},\
  and\ \citenamefont {Farlow}}]{white1989ion}%
  \BibitemOpen
  \bibfield  {author} {\bibinfo {author} {\bibfnamefont {C.}~\bibnamefont
  {White}}, \bibinfo {author} {\bibfnamefont {C.}~\bibnamefont {McHargue}},
  \bibinfo {author} {\bibfnamefont {P.}~\bibnamefont {Sklad}}, \bibinfo
  {author} {\bibfnamefont {L.}~\bibnamefont {Boatner}}, \ and\ \bibinfo
  {author} {\bibfnamefont {G.}~\bibnamefont {Farlow}},\ }\href@noop {}
  {\bibfield  {journal} {\bibinfo  {journal} {Mate. Sci. Rep.}\ }\textbf
  {\bibinfo {volume} {4}},\ \bibinfo {pages} {41} (\bibinfo {year}
  {1989})}\BibitemShut {NoStop}%
\bibitem [{\citenamefont {Yu}\ \emph {et~al.}(1995)\citenamefont {Yu},
  \citenamefont {McIntyre}, \citenamefont {Nastasi},\ and\ \citenamefont
  {Sickafus}}]{yu1995high}%
  \BibitemOpen
  \bibfield  {author} {\bibinfo {author} {\bibfnamefont {N.}~\bibnamefont
  {Yu}}, \bibinfo {author} {\bibfnamefont {P.~C.}\ \bibnamefont {McIntyre}},
  \bibinfo {author} {\bibfnamefont {M.}~\bibnamefont {Nastasi}}, \ and\
  \bibinfo {author} {\bibfnamefont {K.~E.}\ \bibnamefont {Sickafus}},\
  }\href@noop {} {\bibfield  {journal} {\bibinfo  {journal} {Phys. Rev. B}\
  }\textbf {\bibinfo {volume} {52}},\ \bibinfo {pages} {17518} (\bibinfo {year}
  {1995})}\BibitemShut {NoStop}%
\bibitem [{\citenamefont {Pauling}(1935)}]{pauling1935structure}%
  \BibitemOpen
  \bibfield  {author} {\bibinfo {author} {\bibfnamefont {L.}~\bibnamefont
  {Pauling}},\ }\href@noop {} {\bibfield  {journal} {\bibinfo  {journal} {J.
  Am. Chem. Soc.}\ }\textbf {\bibinfo {volume} {57}},\ \bibinfo {pages} {2680}
  (\bibinfo {year} {1935})}\BibitemShut {NoStop}%
\bibitem [{\citenamefont {Kresse}\ and\ \citenamefont
  {Joubert}(1999)}]{kresse1999ultrasoft}%
  \BibitemOpen
  \bibfield  {author} {\bibinfo {author} {\bibfnamefont {G.}~\bibnamefont
  {Kresse}}\ and\ \bibinfo {author} {\bibfnamefont {D.}~\bibnamefont
  {Joubert}},\ }\href@noop {} {\bibfield  {journal} {\bibinfo  {journal} {Phys.
  Rev. B}\ }\textbf {\bibinfo {volume} {59}},\ \bibinfo {pages} {1758}
  (\bibinfo {year} {1999})}\BibitemShut {NoStop}%
\bibitem [{\citenamefont {Bl{\"o}chl}(1994)}]{blochl1994projector}%
  \BibitemOpen
  \bibfield  {author} {\bibinfo {author} {\bibfnamefont {P.~E.}\ \bibnamefont
  {Bl{\"o}chl}},\ }\href@noop {} {\bibfield  {journal} {\bibinfo  {journal}
  {Phys. Rev. B}\ }\textbf {\bibinfo {volume} {50}},\ \bibinfo {pages} {17953}
  (\bibinfo {year} {1994})}\BibitemShut {NoStop}%
\bibitem [{\citenamefont {Kresse}\ and\ \citenamefont
  {Furthm{\"u}ller}(1996{\natexlab{a}})}]{kresse1996efficiency}%
  \BibitemOpen
  \bibfield  {author} {\bibinfo {author} {\bibfnamefont {G.}~\bibnamefont
  {Kresse}}\ and\ \bibinfo {author} {\bibfnamefont {J.}~\bibnamefont
  {Furthm{\"u}ller}},\ }\href@noop {} {\bibfield  {journal} {\bibinfo
  {journal} {Comput. Mater. Sci.}\ }\textbf {\bibinfo {volume} {6}},\ \bibinfo
  {pages} {15} (\bibinfo {year} {1996}{\natexlab{a}})}\BibitemShut {NoStop}%
\bibitem [{\citenamefont {Kresse}\ and\ \citenamefont
  {Furthm{\"u}ller}(1996{\natexlab{b}})}]{kresse1996efficient}%
  \BibitemOpen
  \bibfield  {author} {\bibinfo {author} {\bibfnamefont {G.}~\bibnamefont
  {Kresse}}\ and\ \bibinfo {author} {\bibfnamefont {J.}~\bibnamefont
  {Furthm{\"u}ller}},\ }\href@noop {} {\bibfield  {journal} {\bibinfo
  {journal} {Phys. Rev. B}\ }\textbf {\bibinfo {volume} {54}},\ \bibinfo
  {pages} {11169} (\bibinfo {year} {1996}{\natexlab{b}})}\BibitemShut {NoStop}%
\bibitem [{\citenamefont {Kresse}\ and\ \citenamefont
  {Hafner}(1993)}]{kresse1993ab}%
  \BibitemOpen
  \bibfield  {author} {\bibinfo {author} {\bibfnamefont {G.}~\bibnamefont
  {Kresse}}\ and\ \bibinfo {author} {\bibfnamefont {J.}~\bibnamefont
  {Hafner}},\ }\href@noop {} {\bibfield  {journal} {\bibinfo  {journal} {Phys.
  Rev. B}\ }\textbf {\bibinfo {volume} {48}},\ \bibinfo {pages} {13115}
  (\bibinfo {year} {1993})}\BibitemShut {NoStop}%
\bibitem [{\citenamefont {Perdew}\ \emph {et~al.}(1996)\citenamefont {Perdew},
  \citenamefont {Burke},\ and\ \citenamefont
  {Ernzerhof}}]{perdew1996generalized}%
  \BibitemOpen
  \bibfield  {author} {\bibinfo {author} {\bibfnamefont {J.~P.}\ \bibnamefont
  {Perdew}}, \bibinfo {author} {\bibfnamefont {K.}~\bibnamefont {Burke}}, \
  and\ \bibinfo {author} {\bibfnamefont {M.}~\bibnamefont {Ernzerhof}},\
  }\href@noop {} {\bibfield  {journal} {\bibinfo  {journal} {Phys. Rev. Lett.}\
  }\textbf {\bibinfo {volume} {77}},\ \bibinfo {pages} {3865} (\bibinfo {year}
  {1996})}\BibitemShut {NoStop}%
\bibitem [{\citenamefont {Bilc}\ \emph {et~al.}(2008)\citenamefont {Bilc},
  \citenamefont {Orlando}, \citenamefont {Shaltaf}, \citenamefont {Rignanese},
  \citenamefont {\'I\~niguez},\ and\ \citenamefont
  {Ghosez}}]{PhysRevB.77.165107}%
  \BibitemOpen
  \bibfield  {author} {\bibinfo {author} {\bibfnamefont {D.~I.}\ \bibnamefont
  {Bilc}}, \bibinfo {author} {\bibfnamefont {R.}~\bibnamefont {Orlando}},
  \bibinfo {author} {\bibfnamefont {R.}~\bibnamefont {Shaltaf}}, \bibinfo
  {author} {\bibfnamefont {G.-M.}\ \bibnamefont {Rignanese}}, \bibinfo {author}
  {\bibfnamefont {J.}~\bibnamefont {\'I\~niguez}}, \ and\ \bibinfo {author}
  {\bibfnamefont {P.}~\bibnamefont {Ghosez}},\ }\href@noop {} {\bibfield
  {journal} {\bibinfo  {journal} {Phys. Rev. B}\ }\textbf {\bibinfo {volume}
  {77}},\ \bibinfo {pages} {165107} (\bibinfo {year} {2008})}\BibitemShut
  {NoStop}%
\bibitem [{\citenamefont {Li}\ \emph {et~al.}(2013)\citenamefont {Li},
  \citenamefont {Li}, \citenamefont {Araujo}, \citenamefont {Luo},\ and\
  \citenamefont {Ahuja}}]{C3CY00207A}%
  \BibitemOpen
  \bibfield  {author} {\bibinfo {author} {\bibfnamefont {Y.}~\bibnamefont
  {Li}}, \bibinfo {author} {\bibfnamefont {Y.-L.}\ \bibnamefont {Li}}, \bibinfo
  {author} {\bibfnamefont {C.~M.}\ \bibnamefont {Araujo}}, \bibinfo {author}
  {\bibfnamefont {W.}~\bibnamefont {Luo}}, \ and\ \bibinfo {author}
  {\bibfnamefont {R.}~\bibnamefont {Ahuja}},\ }\href {\doibase
  10.1039/C3CY00207A} {\bibfield  {journal} {\bibinfo  {journal} {Catal. Sci.
  Technol.}\ }\textbf {\bibinfo {volume} {3}},\ \bibinfo {pages} {2214}
  (\bibinfo {year} {2013})}\BibitemShut {NoStop}%
\bibitem [{\citenamefont {Demichelis}\ \emph {et~al.}(2010)\citenamefont
  {Demichelis}, \citenamefont {Civalleri}, \citenamefont {D'Arco},\ and\
  \citenamefont {Dovesi}}]{demichelis2010}%
  \BibitemOpen
  \bibfield  {author} {\bibinfo {author} {\bibfnamefont {R.}~\bibnamefont
  {Demichelis}}, \bibinfo {author} {\bibfnamefont {B.}~\bibnamefont
  {Civalleri}}, \bibinfo {author} {\bibfnamefont {P.}~\bibnamefont {D'Arco}}, \
  and\ \bibinfo {author} {\bibfnamefont {R.}~\bibnamefont {Dovesi}},\
  }\href@noop {} {\bibfield  {journal} {\bibinfo  {journal} {Int. J. Quantum
  Chem}\ }\textbf {\bibinfo {volume} {110}},\ \bibinfo {pages} {2260} (\bibinfo
  {year} {2010})}\BibitemShut {NoStop}%
\bibitem [{\citenamefont {Adamo}\ and\ \citenamefont
  {Barone}(1999)}]{adamo:6158}%
  \BibitemOpen
  \bibfield  {author} {\bibinfo {author} {\bibfnamefont {C.}~\bibnamefont
  {Adamo}}\ and\ \bibinfo {author} {\bibfnamefont {V.}~\bibnamefont {Barone}},\
  }\href@noop {} {\bibfield  {journal} {\bibinfo  {journal} {J. Chem. Phys.}\
  }\textbf {\bibinfo {volume} {110}},\ \bibinfo {pages} {6158} (\bibinfo {year}
  {1999})}\BibitemShut {NoStop}%
\bibitem [{\citenamefont {Guti{\'e}rrez}\ \emph {et~al.}(2001)\citenamefont
  {Guti{\'e}rrez}, \citenamefont {Taga},\ and\ \citenamefont
  {Johansson}}]{gutierrez2001theoretical}%
  \BibitemOpen
  \bibfield  {author} {\bibinfo {author} {\bibfnamefont {G.}~\bibnamefont
  {Guti{\'e}rrez}}, \bibinfo {author} {\bibfnamefont {A.}~\bibnamefont {Taga}},
  \ and\ \bibinfo {author} {\bibfnamefont {B.}~\bibnamefont {Johansson}},\
  }\href@noop {} {\bibfield  {journal} {\bibinfo  {journal} {Phys. Rev. B}\
  }\textbf {\bibinfo {volume} {65}},\ \bibinfo {pages} {012101} (\bibinfo
  {year} {2001})}\BibitemShut {NoStop}%
\bibitem [{\citenamefont {Maglia}\ \emph {et~al.}(2008)\citenamefont {Maglia},
  \citenamefont {Gennari},\ and\ \citenamefont
  {Buscaglia}}]{maglia2008energetics}%
  \BibitemOpen
  \bibfield  {author} {\bibinfo {author} {\bibfnamefont {F.}~\bibnamefont
  {Maglia}}, \bibinfo {author} {\bibfnamefont {S.}~\bibnamefont {Gennari}}, \
  and\ \bibinfo {author} {\bibfnamefont {V.}~\bibnamefont {Buscaglia}},\
  }\href@noop {} {\bibfield  {journal} {\bibinfo  {journal} {J. Am. Ceram.
  Soc.}\ }\textbf {\bibinfo {volume} {91}},\ \bibinfo {pages} {283} (\bibinfo
  {year} {2008})}\BibitemShut {NoStop}%
\bibitem [{\citenamefont {Togo}\ \emph {et~al.}(2008)\citenamefont {Togo},
  \citenamefont {Oba},\ and\ \citenamefont {Tanaka}}]{PhysRevB.78.134106}%
  \BibitemOpen
  \bibfield  {author} {\bibinfo {author} {\bibfnamefont {A.}~\bibnamefont
  {Togo}}, \bibinfo {author} {\bibfnamefont {F.}~\bibnamefont {Oba}}, \ and\
  \bibinfo {author} {\bibfnamefont {I.}~\bibnamefont {Tanaka}},\ }\href@noop {}
  {\bibfield  {journal} {\bibinfo  {journal} {Phys. Rev. B}\ }\textbf {\bibinfo
  {volume} {78}},\ \bibinfo {pages} {134106} (\bibinfo {year}
  {2008})}\BibitemShut {NoStop}%
\bibitem [{\citenamefont {Krokidis}\ \emph {et~al.}(2001)\citenamefont
  {Krokidis}, \citenamefont {Raybaud}, \citenamefont {Gobichon}, \citenamefont
  {Rebours}, \citenamefont {Euzen},\ and\ \citenamefont
  {Toulhoat}}]{krokidis2001theoretical}%
  \BibitemOpen
  \bibfield  {author} {\bibinfo {author} {\bibfnamefont {X.}~\bibnamefont
  {Krokidis}}, \bibinfo {author} {\bibfnamefont {P.}~\bibnamefont {Raybaud}},
  \bibinfo {author} {\bibfnamefont {A.-E.}\ \bibnamefont {Gobichon}}, \bibinfo
  {author} {\bibfnamefont {B.}~\bibnamefont {Rebours}}, \bibinfo {author}
  {\bibfnamefont {P.}~\bibnamefont {Euzen}}, \ and\ \bibinfo {author}
  {\bibfnamefont {H.}~\bibnamefont {Toulhoat}},\ }\href@noop {} {\bibfield
  {journal} {\bibinfo  {journal} {J. Phys. Chem. B}\ }\textbf {\bibinfo
  {volume} {105}},\ \bibinfo {pages} {5121} (\bibinfo {year}
  {2001})}\BibitemShut {NoStop}%
\bibitem [{\citenamefont {Pinto}\ \emph {et~al.}(2004)\citenamefont {Pinto},
  \citenamefont {Nieminen},\ and\ \citenamefont
  {Elliott}}]{PhysRevB.70.125402}%
  \BibitemOpen
  \bibfield  {author} {\bibinfo {author} {\bibfnamefont {H.~P.}\ \bibnamefont
  {Pinto}}, \bibinfo {author} {\bibfnamefont {R.~M.}\ \bibnamefont {Nieminen}},
  \ and\ \bibinfo {author} {\bibfnamefont {S.~D.}\ \bibnamefont {Elliott}},\
  }\href {\doibase 10.1103/PhysRevB.70.125402} {\bibfield  {journal} {\bibinfo
  {journal} {Phys. Rev. B}\ }\textbf {\bibinfo {volume} {70}},\ \bibinfo
  {pages} {125402} (\bibinfo {year} {2004})}\BibitemShut {NoStop}%
\bibitem [{\citenamefont {Ealet}\ \emph {et~al.}(1994)\citenamefont {Ealet},
  \citenamefont {Elyakhloufi}, \citenamefont {Gillet},\ and\ \citenamefont
  {Ricci}}]{ealet1994electronic}%
  \BibitemOpen
  \bibfield  {author} {\bibinfo {author} {\bibfnamefont {B.}~\bibnamefont
  {Ealet}}, \bibinfo {author} {\bibfnamefont {M.}~\bibnamefont {Elyakhloufi}},
  \bibinfo {author} {\bibfnamefont {E.}~\bibnamefont {Gillet}}, \ and\ \bibinfo
  {author} {\bibfnamefont {M.}~\bibnamefont {Ricci}},\ }\href@noop {}
  {\bibfield  {journal} {\bibinfo  {journal} {Thin Solid Films}\ }\textbf
  {\bibinfo {volume} {250}},\ \bibinfo {pages} {92} (\bibinfo {year}
  {1994})}\BibitemShut {NoStop}%
\bibitem [{\citenamefont {Kefi}\ \emph {et~al.}(1993)\citenamefont {Kefi},
  \citenamefont {Jonnard}, \citenamefont {Vergand}, \citenamefont {Bonnelle},\
  and\ \citenamefont {Gillet}}]{kefi1993hybridization}%
  \BibitemOpen
  \bibfield  {author} {\bibinfo {author} {\bibfnamefont {M.}~\bibnamefont
  {Kefi}}, \bibinfo {author} {\bibfnamefont {P.}~\bibnamefont {Jonnard}},
  \bibinfo {author} {\bibfnamefont {F.}~\bibnamefont {Vergand}}, \bibinfo
  {author} {\bibfnamefont {C.}~\bibnamefont {Bonnelle}}, \ and\ \bibinfo
  {author} {\bibfnamefont {E.}~\bibnamefont {Gillet}},\ }\href@noop {}
  {\bibfield  {journal} {\bibinfo  {journal} {J. Phys.: Condens. Matter}\
  }\textbf {\bibinfo {volume} {5}},\ \bibinfo {pages} {8629} (\bibinfo {year}
  {1993})}\BibitemShut {NoStop}%
\bibitem [{\citenamefont {Snijders}\ \emph {et~al.}(2002)\citenamefont
  {Snijders}, \citenamefont {Jeurgens},\ and\ \citenamefont
  {Sloof}}]{snijders2002structure}%
  \BibitemOpen
  \bibfield  {author} {\bibinfo {author} {\bibfnamefont {P.}~\bibnamefont
  {Snijders}}, \bibinfo {author} {\bibfnamefont {L.}~\bibnamefont {Jeurgens}},
  \ and\ \bibinfo {author} {\bibfnamefont {W.}~\bibnamefont {Sloof}},\
  }\href@noop {} {\bibfield  {journal} {\bibinfo  {journal} {Surf. Sci.}\
  }\textbf {\bibinfo {volume} {496}},\ \bibinfo {pages} {97} (\bibinfo {year}
  {2002})}\BibitemShut {NoStop}%
\bibitem [{\citenamefont {Christensen}\ \emph {et~al.}(1982)\citenamefont
  {Christensen}, \citenamefont {Lehmann},\ and\ \citenamefont
  {Convert}}]{christensen1982}%
  \BibitemOpen
  \bibfield  {author} {\bibinfo {author} {\bibfnamefont {A.~N.}\ \bibnamefont
  {Christensen}}, \bibinfo {author} {\bibfnamefont {M.}~\bibnamefont
  {Lehmann}}, \ and\ \bibinfo {author} {\bibfnamefont {P.}~\bibnamefont
  {Convert}},\ }\href@noop {} {\bibfield  {journal} {\bibinfo  {journal} {Acta
  Chem. Scand. A}\ }\textbf {\bibinfo {volume} {36}} (\bibinfo {year}
  {1982})}\BibitemShut {NoStop}%
\bibitem [{\citenamefont {Loyola}\ \emph {et~al.}(2010)\citenamefont {Loyola},
  \citenamefont {Men{\'e}ndez-Proupin},\ and\ \citenamefont
  {Guti{\'e}rrez}}]{loyola2010atomistic}%
  \BibitemOpen
  \bibfield  {author} {\bibinfo {author} {\bibfnamefont {C.}~\bibnamefont
  {Loyola}}, \bibinfo {author} {\bibfnamefont {E.}~\bibnamefont
  {Men{\'e}ndez-Proupin}}, \ and\ \bibinfo {author} {\bibfnamefont
  {G.}~\bibnamefont {Guti{\'e}rrez}},\ }\href@noop {} {\bibfield  {journal}
  {\bibinfo  {journal} {J. Mater. Sci.}\ }\textbf {\bibinfo {volume} {45}},\
  \bibinfo {pages} {5094} (\bibinfo {year} {2010})}\BibitemShut {NoStop}%
\bibitem [{\citenamefont {Roy}\ and\ \citenamefont
  {Sood}(1995)}]{roy1995phonons}%
  \BibitemOpen
  \bibfield  {author} {\bibinfo {author} {\bibfnamefont {A.}~\bibnamefont
  {Roy}}\ and\ \bibinfo {author} {\bibfnamefont {A.~K.}\ \bibnamefont {Sood}},\
  }\href@noop {} {\bibfield  {journal} {\bibinfo  {journal} {Pramana}\ }\textbf
  {\bibinfo {volume} {44}},\ \bibinfo {pages} {201} (\bibinfo {year}
  {1995})}\BibitemShut {NoStop}%
\bibitem [{\citenamefont {Ruan}\ \emph {et~al.}(2001)\citenamefont {Ruan},
  \citenamefont {Frost},\ and\ \citenamefont {Kloprogge}}]{ruan2001comparison}%
  \BibitemOpen
  \bibfield  {author} {\bibinfo {author} {\bibfnamefont {H.}~\bibnamefont
  {Ruan}}, \bibinfo {author} {\bibfnamefont {R.}~\bibnamefont {Frost}}, \ and\
  \bibinfo {author} {\bibfnamefont {J.}~\bibnamefont {Kloprogge}},\ }\href@noop
  {} {\bibfield  {journal} {\bibinfo  {journal} {J. Raman Spectrosc.}\ }\textbf
  {\bibinfo {volume} {32}},\ \bibinfo {pages} {745} (\bibinfo {year}
  {2001})}\BibitemShut {NoStop}%
\bibitem [{\citenamefont {Aminzadeh}\ and\ \citenamefont
  {Sarikhani-Fard}(1999)}]{aminzadeh1999raman}%
  \BibitemOpen
  \bibfield  {author} {\bibinfo {author} {\bibfnamefont {A.}~\bibnamefont
  {Aminzadeh}}\ and\ \bibinfo {author} {\bibfnamefont {H.}~\bibnamefont
  {Sarikhani-Fard}},\ }\href@noop {} {\bibfield  {journal} {\bibinfo  {journal}
  {Spectrochim. Acta, Part A}\ }\textbf {\bibinfo {volume} {55}},\ \bibinfo
  {pages} {1421} (\bibinfo {year} {1999})}\BibitemShut {NoStop}%
\bibitem [{\citenamefont {Otto}\ \emph {et~al.}(1992)\citenamefont {Otto},
  \citenamefont {Hubbard}, \citenamefont {Weber},\ and\ \citenamefont
  {Graham}}]{otto1992raman}%
  \BibitemOpen
  \bibfield  {author} {\bibinfo {author} {\bibfnamefont {K.}~\bibnamefont
  {Otto}}, \bibinfo {author} {\bibfnamefont {C.}~\bibnamefont {Hubbard}},
  \bibinfo {author} {\bibfnamefont {W.}~\bibnamefont {Weber}}, \ and\ \bibinfo
  {author} {\bibfnamefont {G.}~\bibnamefont {Graham}},\ }\href@noop {}
  {\bibfield  {journal} {\bibinfo  {journal} {Appl. Catal., B}\ }\textbf
  {\bibinfo {volume} {1}},\ \bibinfo {pages} {317} (\bibinfo {year}
  {1992})}\BibitemShut {NoStop}%
\bibitem [{\citenamefont {Dyer}\ \emph {et~al.}(1993)\citenamefont {Dyer},
  \citenamefont {Hendra}, \citenamefont {Forsling},\ and\ \citenamefont
  {Ranheimer}}]{dyer1993surface}%
  \BibitemOpen
  \bibfield  {author} {\bibinfo {author} {\bibfnamefont {C.}~\bibnamefont
  {Dyer}}, \bibinfo {author} {\bibfnamefont {P.~J.}\ \bibnamefont {Hendra}},
  \bibinfo {author} {\bibfnamefont {W.}~\bibnamefont {Forsling}}, \ and\
  \bibinfo {author} {\bibfnamefont {M.}~\bibnamefont {Ranheimer}},\ }\href@noop
  {} {\bibfield  {journal} {\bibinfo  {journal} {Spectrochim. Acta, Part A}\
  }\textbf {\bibinfo {volume} {49}},\ \bibinfo {pages} {691} (\bibinfo {year}
  {1993})}\BibitemShut {NoStop}%
\bibitem [{mal()}]{mallard1992nist}%
  \BibitemOpen
  \href@noop {} {\emph {\bibinfo {title}
  {http://www.nist.gov/srd/onlinelist.cfm}}}\BibitemShut {NoStop}%
\bibitem [{alt()}]{alteo}%
  \BibitemOpen
  \href@noop {} {\emph {\bibinfo {title}
  {http://www.alteo-alumina.com/en/transition-aluminas}}}\BibitemShut {NoStop}%
\bibitem [{\citenamefont {Digne}\ \emph {et~al.}(2002)\citenamefont {Digne},
  \citenamefont {Sautet}, \citenamefont {Raybaud}, \citenamefont {Euzen},\ and\
  \citenamefont {Toulhoat}}]{digne2002hydroxyl}%
  \BibitemOpen
  \bibfield  {author} {\bibinfo {author} {\bibfnamefont {M.}~\bibnamefont
  {Digne}}, \bibinfo {author} {\bibfnamefont {P.}~\bibnamefont {Sautet}},
  \bibinfo {author} {\bibfnamefont {P.}~\bibnamefont {Raybaud}}, \bibinfo
  {author} {\bibfnamefont {P.}~\bibnamefont {Euzen}}, \ and\ \bibinfo {author}
  {\bibfnamefont {H.}~\bibnamefont {Toulhoat}},\ }\href@noop {} {\bibfield
  {journal} {\bibinfo  {journal} {J. Catal.}\ }\textbf {\bibinfo {volume}
  {211}},\ \bibinfo {pages} {1} (\bibinfo {year} {2002})}\BibitemShut {NoStop}%
\bibitem [{\citenamefont {Tops{\o}e}\ \emph {et~al.}(2004)\citenamefont
  {Tops{\o}e}, \citenamefont {N{\o}rskov} \emph {et~al.}}]{topsoe2004negative}%
  \BibitemOpen
  \bibfield  {author} {\bibinfo {author} {\bibfnamefont {N.-Y.}\ \bibnamefont
  {Tops{\o}e}}, \bibinfo {author} {\bibfnamefont {J.~K.}\ \bibnamefont
  {N{\o}rskov}},  \emph {et~al.},\ }\href@noop {} {\bibfield  {journal}
  {\bibinfo  {journal} {Nature materials}\ }\textbf {\bibinfo {volume} {3}},\
  \bibinfo {pages} {289} (\bibinfo {year} {2004})}\BibitemShut {NoStop}%
\bibitem [{\citenamefont {Klug}\ \emph {et~al.}(1987)\citenamefont {Klug},
  \citenamefont {Prochazka},\ and\ \citenamefont {Doremus}}]{klug1987alumina}%
  \BibitemOpen
  \bibfield  {author} {\bibinfo {author} {\bibfnamefont {F.~J.}\ \bibnamefont
  {Klug}}, \bibinfo {author} {\bibfnamefont {S.}~\bibnamefont {Prochazka}}, \
  and\ \bibinfo {author} {\bibfnamefont {R.~H.}\ \bibnamefont {Doremus}},\
  }\href@noop {} {\bibfield  {journal} {\bibinfo  {journal} {J. Am. Ceram.
  Soc.}\ }\textbf {\bibinfo {volume} {70}},\ \bibinfo {pages} {750} (\bibinfo
  {year} {1987})}\BibitemShut {NoStop}%
\bibitem [{\citenamefont {Kemp}\ and\ \citenamefont
  {Pitzer}(1937)}]{kemp1937entropy}%
  \BibitemOpen
  \bibfield  {author} {\bibinfo {author} {\bibfnamefont {J.}~\bibnamefont
  {Kemp}}\ and\ \bibinfo {author} {\bibfnamefont {K.~S.}\ \bibnamefont
  {Pitzer}},\ }\href@noop {} {\bibfield  {journal} {\bibinfo  {journal} {J. Am.
  Chem. Soc.}\ }\textbf {\bibinfo {volume} {59}},\ \bibinfo {pages} {276}
  (\bibinfo {year} {1937})}\BibitemShut {NoStop}%
\end{thebibliography}

\end{document}